\documentstyle [eepic,epic,epsf] {article}

\title{Deriving Abstract Semantics for Forward Analysis 
       of Normal Logic Programs~\footnote{This draft is a reformulation of Chapter 4 in~\cite{LU94}}}
\author{Lunjin Lu\\ Department of Computer Science\\ 
The University of Waikato\\ Hamilton, New Zealand} 
\date{}

\begin{document}
\bibliographystyle{plain}

%\pagestyle{empty}
%\tableofcontents

\maketitle

\newtheorem {definition} {Definition} [section]
\newtheorem {hypothesis}[definition]{Hypothesis}
\newtheorem {proposition}[definition]{Proposition}
\newtheorem {corollary}[definition]{Corollary}
\newtheorem {lemma}[definition]{Lemma}
\newtheorem {observation}[definition]{Observation}
\newtheorem {theorem}[definition]{Theorem}
\newtheorem {transformation}[definition]{Transformation}
\newtheorem {algorithm}[definition]{Algorithm}
\newtheorem {program}[definition]{Program}
\newtheorem {example}[definition]{Example}

\newcommand {\comments} [1]{}
\newcommand {\definedas}{\stackrel{\em def}{=}}
\newcommand {\fcomp}{\cdot}
\newcommand {\values}{/}
\newcommand {\first}[1]{{\it #1}}
\newcommand {\Vect}[1]{\mbox{$#1_1,#1_2,\cdots,#1_n$}}
\newcommand {\Vector}[2]{\mbox{$#1_1,\cdots,#1_#2$}}

\newcommand{\circled}[1] {\mbox{$\bigcirc$\hspace{-1em}
      \raisebox{0.2ex}{$\scriptstyle #1$}\hspace{.4ex}}}
\newcommand{\point}[1] {\circled{#1}}
\newcommand{\edge}[2]{{#1
      \raisebox{-.2ex}{$\leftarrow\!\!\!\!\bullet$}#2}}
\newcommand{\stackitem}[3]{\mbox{$\parallel\edge{#1}{#2},#3\parallel$}}

\def\term{{\sf TERM}}
\def\atom{{\sf ATOM}}
\def\syneq{\equiv}

\def\thus{\hbox to 12pt{.\raise 4pt \hbox{.}.}}

\def\therefore{\mbox{$\mathrel{.\dot{~}.}$~}}
\def\because{\mbox{$\mathrel{\dot{~}.\dot{~}}$~}}

\def\functional{\mbox{\raisebox{0.2ex}{$\scriptstyle\mapsto$}\hspace{-0.9em}$\bigcirc$}}
\def\free{\circled{f}}
\def\ground{\circled{g}}
\def\other{\circled{o}} 

\def\transfers{\stackrel{P}{\leadsto}}

\def\allvars{\mbox{$\cal V\!\!A\!R$}}

\def\dcal{{\cal D}}
\def\scal{{\cal S}}
\def\lcal{{\cal L}}
\def\vcal{{\cal V}}
\def\wcal{{\cal W}}
\def\gcal{{\cal G}}
\def\ncal{{\cal N}}
\def\ecal{{\cal E}}
\def\ucal{{\cal U}}
\def\ycal{{\cal Y}}
\def\zcal{{\cal Z}}

\def\fail{{\sf fail}}

\newenvironment{proof}{{\sc Proof:\/}\em }{~\hspace*{\fill}
  $\rule{1ex}{1.5ex}$\par}

\newcommand{\proofoflemma}[2] {\underline{{\sc Proof of
      lemma}~\ref{#1} (p.\pageref{#1})}:\/ #2 ~\hspace*{\fill}
  $\rule{1ex}{1.5ex}$\par \vspace{1pc}}

\newcommand{\proofoftheorem}[2] {\underline{{\sc Proof of
      theorem}~\ref{#1} (p.\pageref{#1})}:\/ #2 ~\hspace*{\fill}
  $\rule{1ex}{1.5ex}$\par\vspace{1pc}}

\def\fsem{F^\flat_{P}}
\def\fdom{\dcal^\flat}
\def\fgamma{\gamma^\flat} 
\def\fbigx{X^\flat}
\def\fbigy{Y^\flat}
\def\fbeta{\beta^\flat}
\def\feta{\eta^\flat}
\def\fsigma{\sigma^\flat}
\def\ftheta{\theta^\flat}
\def\fzeta{\zeta^\flat}
\def\fsqsubseteq{\sqsubseteq^\flat}
\def\fsqcap{\sqcap^\flat}
\def\fsqcup{\sqcup^\flat}
\def\ftop{\top^\flat}
\def\fbot{\bot^\flat}

\def\dsem{F^\diamond_{P}}
\def\ddom{\dcal^\diamond}
\def\dgamma{\gamma^\diamond}
\def\dsigma{\sigma^\diamond}
\def\dtheta{\theta^\diamond}
\def\dbigx{X^\diamond}
\def\dbigy{Y^\diamond}
\def\dsqsubseteq{\sqsubseteq^\diamond}
\def\dsqcap{\sqcap^\diamond}
\def\dsqcup{\sqcup^\diamond}
\def\dtop{\top^\diamond}
\def\dbot{\bot^\diamond}

\def\ssem{F^\sharp_{P}}
\def\sdom{\dcal^\sharp}
\def\sgamma{\gamma^\sharp}
\def\sbigx{X^\sharp}
\def\sbigy{Y^\sharp}
\def\ssqsubseteq{\sqsubseteq^\sharp}
\def\ssqcap{\sqcap^\sharp}
\def\ssqcup{\sqcup^\sharp}
\def\stop{\top^\sharp}
\def\sbot{\bot^\sharp}
\def\sunify{unify^\sharp}
\def\stackitems{\scal^\sharp}

\def\sem{F_{P}}
\def\dom{\dcal}
\def\unify{unify}

\def\initstacks{{\scal}_{0}}
\def\finalstacks{{\scal}_{\infty}}
\def\stacks{{\scal}}

\def\edges{\ecal_{P}}
\def\points{\ncal_{P}}
\def\lfp{{\em lfp}}

\newcommand{\asub}[1]   {\overline{ASub}_{#1}}
\newcommand{\asubgamma}[1] {\overline{\gamma}_{#1}}
\newcommand{\asuborder}[1] {\overline{\sqsubseteq}_{#1}}
\newcommand{\asubbot}[1]  {\overline{\bot}_{#1}}
\newcommand{\asubtop}[1]  {\overline{\top}_{#1}}
\newcommand{\asubcup}[1]  {\overline{\sqcup}_{#1}}
\newcommand{\asubcap}[1]  {\overline{\sqcap}_{#1}}
\newcommand{\aunify}[2]   {\overline{unify}_{#1,#2}}
\newcommand{\asubid}[1] {\overline{\epsilon}_{#1}}

\def\asubf {\overline{ASub}}
\def\asubgammaf{\overline{\gamma}}
\def\asubcupf{\overline{\sqcup}}
\def\asubidf{\overline{\epsilon}}
\def\aunifyf{\overline{unify}}
\def\dunifyf{\widehat{unify}}

\begin{abstract}
  The problem of forward abstract interpretation of {\em normal} logic
  programs has not been formally addressed in the literature although
  negation as failure is dealt with through the built-in predicate ${!}$ in
  the way it is implemented in Prolog. This paper proposes a solution to
  this problem by deriving two generic fixed-point abstract semantics
  $\fsem$ and $\dsem$ for forward abstract interpretation of {\em normal}
  logic programs.

  $\fsem$ is intended for inferring data descriptions for edges in the
  program graph where an edge denotes the possibility that the control of
  execution transfers from its source program point to its destination
  program point. $\dsem$ is derived from $\fsem$ and is intended for
  inferring data descriptions for textual program points.

\end{abstract}

\section {Introduction}

Abstract interpretation~\cite{Cousot:Cousot:77} is a program analysis
methodology for statically deriving run-time properties of programs.
The derived program properties are then used by other program
processors such as compilers, partial evaluators, etc.  Program
analyses are viewed as program executions over non-standard data
domains.  Cousot and Cousot first laid solid mathematical foundations
for abstract interpretation~\cite{Cousot:Cousot:77,Cousot:Cousot:79}.
The idea is to define a collecting semantics for a program which
associates with each program point the set of the storage states that
are obtained whenever the execution reaches the point.  Then an
approximation of the collecting semantics is calculated by simulating
over a non-standard data domain the computation of the collecting
semantics over the standard data domain. The standard data domain is
called the concrete domain and the non-standard domain is called the
abstract domain. 

There has been recently much research into abstract interpretation of
logic programs~\cite{Cousot:JLP92}.  Abstract interpretation has been
used in both forward and backward analyses of logic programs.  A
forward analysis~\cite{Bruynooghe91} approximates the set of
substitutions that might occur at a program point given a program and
a set of goal descriptions. A backward
analysis~\cite{BarbutiGL93,CodishDY94,Marriott:Sondergaard:88,Marriott:JLP92}
approximates the set of the atoms that are logical consequences of a
program~\cite{Emden:Kowalski:76}.  A number of generic abstract
semantics, often called frameworks,
schemes~\cite{Bruynooghe91,Kanamori:93,Mellish:87}, have been proposed
for forward abstract interpretation of logic programs. These generic
abstract semantics have been specialised for the detection of
determinacy~\cite{Debray89b}, data dependency
analyses~\cite{CodishDY91,Debray89a,Jacobs89:NACLP,Jacobs:JLP92,MuthukumarH91,Muthukumar:JLP92},
mode
inference~\cite{BruynoogheJCD87,CodishDY91,Debray89a,MarienJMB89,MuthukumarH91,Taylor89},
program transformation~\cite{Schreye:Bruynooghe:88}, type
inference~\cite{BruynoogheJCD87,HoriuchiK87,Janssens:JLP92,MarienJMB89},
termination proof~\cite{VerschaetseS91}, etc. However, these generic
abstract semantics have been developed for forward abstract
interpretation of {\em definite} logic programs. The problem of
forward abstract interpretation of {\em normal} logic programs has not
been formally addressed in the literature although negation as failure
is dealt with in practice through the built-in predicate ${!}$ in the
way it is implemented in Prolog. This paper proposes a solution to
this problem by deriving two generic fixed-point abstract semantics
$\fsem$ and $\dsem$ for forward abstract interpretation of {\em
normal} logic programs without relying on any capability of dealing
with the built-in predicate ${!}$.

  Existing generic abstract semantics in the literature are
  optimisation-oriented and they are used to infer data descriptions
  for textual program points. However, there are some applications
  such as debugging where it is helpful to infer data descriptions for
  edges $\edge{p}{q}$ in the program graph where $\edge{p}{q}$ denotes
  that the control of execution may transfers from program point $q$
  to program point $p$.  $\fsem$ is intended for these
  applications. $\dsem$ is derived from $\fsem$ and is
  intended for inferring data descriptions for textual program points.

  The way in which $\fsem$ and $\dsem$ are derived is
  conventional. $\fsem$ is based on a fixed-point collecting
  semantics that associates a set of substitutions with each edge
  $\edge{p}{q}$ in the program graph. The set of substitutions
  associated with $\edge{p}{q}$ includes all the substitutions at
  program point $p$ whenever the control of execution transfers from
  program point $q$ to program point $p$.  The collecting semantics is
  obtained through two approximations.  The operational semantics
  SLDNF-resolution via the left-to-right computation rule is first
  approximated by a transition system.  The transition system is then
  approximated by the collecting semantics.  Obtained from the
  collecting semantics by a further approximation is $\fsem$
  that can then be specialised to perform various analyses
  under certain sufficient conditions. $\dsem$ is derived
  from $\fsem$ through  one more approximation and it can be
  specialised to perform various analyses under the same
  sufficient conditions as $\fsem$.

  This paper makes two contributions. Firstly, $\fsem$ and $\dsem$ can
  be used to perform forward abstract interpretation of {\em normal}
  logic programs while the existing generic abstract semantics in the
  literature can only be used to perform forward abstract
  interpretation of {\em definite} logic programs. A common practice
  in analysing logic programs with negation as failure is to extend an
  existing generic abstract semantics with capability of dealing with
  the built-in predicate $!$ and then analyse negation as failure in
  the way it is implemented in Prolog. However, the built-in predicate
  $!$ is not a part of the language of normal logic programs. The
  derivation of $\fsem$ and $\dsem$ does not resort to any capability
  of dealing with any built-in predicate.  Secondly, $\fsem$ is easier
  to specialise for inferring data descriptions for edges in the
  program graph than the existing generic abstract semantics in the
  literature.

  The remainder of this paper is organised as follows.
  Section~\ref{scheme:section:preliminaries} briefly recalls on
  mathematical foundations for abstract interpretation and some
  terminology in logic program, and introduces some notations used
  later in this paper. Section~\ref{scheme:section:operational}
  reformulates SLDNF in order to facilitate the derivation of the
  collecting semantics.  Section~\ref{scheme:section:static} derives
  the collecting semantics from the operational semantics.
  Section~\ref{scheme:section:abstract} derives $\fsem$ from the
  collecting semantics through a further approximation, and gives the
  sufficient conditions for $\fsem$ to safely approximate the
  collecting semantics, and analyses its worst case complexity.
  Section~\ref{scheme:sec:simpler} derives $\dsem$ from
  $\fsem$ through one more approximation, and analyses its worst
  case complexity.  In section~\ref{scheme:section:example}, we show
  how $\fsem$ and $\dsem$ can be specialised to infer
  groundness information.  Section~\ref{scheme:section:discusion}
  reviews related work on forward abstract interpretation of logic
  programs.
  Section~\ref{scheme:section:conclusion} concludes the paper.

\section {Preliminaries} \label{scheme:section:preliminaries}
\subsection{Complete lattice} 

Let $S,S_{1},S_{2}$ be sets. The \first{powerset} $\wp(S)$ of $S$ is
the set of subsets of $S$.  $\wp(S)=\{X~|~X\subseteq S\}$. The
\first{Cartesian product} $S_{1}\times S_{2}$ of $S_{1}$ and $S_{2}$
is the set of the tuples with the first components in $S_{1}$ and
the second components in $S_{2}$.  \mbox{$S_{1}\times
  S_{2}=\{<s_{1},s_{2}>~|~s_{1}\in S_{1}\wedge s_{2}\in S_{2}\}$}.

A \first{binary relation} $R$ on $S$ is a subset of $S\times S$.
\mbox{$<x,y>\in R$} is denoted as $xRy$ and \mbox{$<x,y>\not\in R$} is
denoted as \mbox{$x\!\!\not\!\!Ry$}. $R$ is \first{reflexive} iff
$xRx$ for every $x\in S$.  $R$ is \first{transitive} iff for every
$x,y,z\in S$, $xRy$ and $yRz$ implies $xRz$.  $R$ is
\first{anti-symmetric} if, for every $x,y\in S$, $xRy$ and $yRx$
implies $x=y$. 

A \first{partial order} $\sqsubseteq$ on $S$ is a reflexive,
anti-symmetric and transitive relation on $S$. A \first{poset}
$<S,\sqsubseteq>$ is tuple where $S$ is a set and $\sqsubseteq$ is a
partial order on $S$. 

Let $<S,\sqsubseteq>$ be a poset, $X\subseteq S$ and $u,v\in S$.  $u$
is an \first{upper bound} of $X$ if $x\sqsubseteq u$ for every $x\in
X$.  An upper bound $u$ of $X$ is the \first{least upper bound} of $X$
if $u\sqsubseteq v$ for every other upper bound $v$ of $X$. The least
upper bound of $X$ is unique if it exists and is denoted as $\sqcup
X$.  $\sqcup \{x_{1},x_{2},\ldots,x_{k}\}$ is sometimes written as
$x_{1} \sqcup x_{2} \sqcup \cdots \sqcup x_{k}$. $\sqcup \{x|P(x)\}$
is sometimes written as $\sqcup_{P(x)}x$.  Similarly, $u$ is a
\first{lower bound} of $X$ if $u\sqsubseteq x$ for every $x\in X$. A
lower bound $u$ of $X$ is the \first{greatest lower  bound} of $X$ if
$v\sqsubseteq u$ for every other lower bound $v$ of $X$. The greatest
lower bound of $X$ is unique if it exists and is denoted as $\sqcap
X$. $\sqcap \{x_{1},x_{2},\ldots,x_{k}\}$ is sometimes written as
$x_{1} \sqcap x_{2} \sqcap \cdots \sqcap x_{k}$.  $\sqcap \{x|P(x)\}$
is sometimes written as $\sqcap_{P(x)}x$.

Let $<S,\sqsubseteq>$ be a poset. $\bot \in S$ is an \first{infimum}
of $<S,\sqsubseteq>$ if $\bot\sqsubseteq x$ for every $x\in S$.  Not
every poset has an infimum. A poset has a unique infimum when it has
one. A \first{supremum} $\top$ of $<S,\sqsubseteq>$ is defined dually.

A \first{complete lattice} $<S,\sqsubseteq>$ is a poset such that
every $X\subseteq S$ has a least upper bound and a greatest lower
bound. A complete lattice has a unique infimum and a unique supremum.
We will write a complete lattice $<S,\sqsubseteq>$ as
$<S,\sqsubseteq,\bot,\top,\sqcap,\sqcup>$ when it is necessary to make
the infimum $\bot$, the supremum $\top$, the greatest lower bound
operator $\sqcap$ and the lease upper bound operator $\sqcup$
explicit.

Let $D$ and $\bar{D}$ be sets. $D\mapsto \bar{D}$ denotes the set of
total functions from $D$ to $\bar{D}$.  A \first{total function} $f$
from $D$ to $\bar{D}$ is a subset of $D\times \bar{D}$ such that, for
every $d\in D$, there is one and only one $\bar{d}\in \bar{D}$ such
that \mbox{$<d,\bar{d}>\in f$}.  \mbox{$<d,\bar{d}>\in f$} is denoted
as $\bar{d}=f(d)$. 

Let $f\in D\mapsto\bar{D}$ and $g\in\bar{D}\mapsto\tilde{D}$. We use
$g\fcomp f$ to denote the \first{composition of two functions} $f$ and
$g$. $g\fcomp f \definedas \lambda x\in D. g(f(x))$.

Let $<D,\sqsubseteq,\bot,\top,\sqcap,\sqcup>$ and
$<\bar{D},\bar{\sqsubseteq},\bar{\bot},
\bar{\top},\bar{\sqcap},\bar{\sqcup}>$ be complete lattices, and $f\in
D \mapsto \bar{D}$.  $f$ is \first{monotonic} if
$f(x)~\bar{\sqsubseteq}~f(y)$ for any $x,y\in
D$ such that ${x}~{\sqsubseteq}~{y}$. 

Let $<D,\sqsubseteq>$ be a complete lattice and $f\in D\mapsto D$.
$x\in D$ is a \first{fixed-point} of $f$ if $x=f(x)$. $x$ is the
\first{least fixed-point}, denoted by $\lfp f$, of $f$ if
$x~\sqsubseteq~y$ for each fixed-point $y$ of $f$.  
$\lfp f=f\uparrow \beta$ for some
ordinal $\beta$~\cite{Tarski55} where
\[
  f\uparrow \beta \definedas \left\{
        \begin{array}{lr}
            \sqcup \{ f\uparrow \beta'~|~\beta' < \beta \} &
            if~\beta~is~a~limit~ordinal\\
            f(f\uparrow (\beta-1)) & if~\beta~is~a~successor~ordinal
        \end{array} 
  \right.
\]

\subsection {Abstract Interpretation} 
We now formalise the notion of abstract interpretation according to
the ideas given by~\cite{Cousot:Cousot:77}. The idea of having an
element in an abstract domain $<\bar{D},\bar{\sqsubseteq}>$ as
description of a element in a concrete domain $<D,\sqsubseteq>$ is
formalised by a monotonic function from $\bar{D}$ to $D$, called a
\first{concretisation function}.

We say that an element $d$ in $D$ is \first{approximated} by an
element $\bar{d}$ in $\bar{D}$ if
$d~\bar{\sqsubseteq}~\gamma(\bar{d})$. There might well be a number of
elements that approximate $d$.  If $\gamma\in \bar{D}\mapsto D$ and
$\gamma'\in \tilde{D}\mapsto \bar{D}$ are concretisation functions
then $\gamma\fcomp\gamma'\in \tilde{D}\mapsto D$ is a concretisation
function.

The notion of approximation can also be formalised by means of 
an abstraction function from the concrete domain to the abstract
domain, or a Galois connection between  the abstract domain and the
concrete domain or a relation between the concrete and the abstract
domains~\cite{Marriott93}.

A fixed-point interpretation of a program is the least fixed-point of
a function associated with the program on a semantic domain, often a
complete lattice.  The following theorem shows how the least
fixed-point of a monotonic function on one complete lattice can be
approximated by the least fixed-point of another monotonic function
on another complete lattice.
\begin{theorem} \label{frame:th:cousot} {\em 
    If \mbox{$<D,\sqsubseteq>$} and
    \mbox{$<\bar{D},\bar{\sqsubseteq}>$} are complete lattices, $F$ a
    monotonic function on $<D,\sqsubseteq>$, $\bar{F}$ a monotonic
    function on $<\bar{D},\bar{\sqsubseteq}>$, $\gamma$ a monotonic function
    from $\bar{D}$ to $D$ and\linebreak $\forall \bar{d}\in \bar{D}.(
    F \fcomp \gamma(\bar{d}) ~\sqsubseteq~ \gamma \fcomp
    \bar{F}(\bar{d}))$ then $\lfp F ~\sqsubseteq~ \gamma(\lfp
    \bar{F})$.}

\begin{proof} See~\cite{Marriott:JLP92}. \end{proof}
\end{theorem} 

\subsection {Logic programming} \label{pre:sec:lp}

We assume that the reader is familiar with the terminology in logic
programming~\cite{Lloyd:87}.  Let $\lcal$ be a first order language
with function symbol set $\Sigma$ and predicate symbol set $\Pi$ which
is disjoint from $\Sigma$.  Let $\allvars$ be a denumerable set of
variables and $\vcal\subseteq \allvars$. $\term(\Sigma,\vcal)$ denotes
the set of \first{terms} that can be built from $\Sigma$ and $\vcal$.
  $\atom(\Pi,\Sigma,\vcal)$ denotes the set of \first{atoms}
  constructible from $\Pi$, $\Sigma$ and $\vcal$.  The negation of an
  atom $A$ is denoted as $\neg A$. A \first{literal} is either an atom
  or the negation of an atom.

  Let $\theta$ and $\sigma$ be \first{substitutions}.
  $\sigma\circ\theta$ denotes the composition of $\sigma$ and
  $\theta$. $dom(\theta)$ denotes the \first{domain} of $\theta$.
  Define $Sub\definedas \{\theta~|~\theta~is~a~substitution\}$. An
  \first{expression} is a term, an atom, a literal, a clause, a goal
  etc. The set of variables in an expression $E$ is denoted as
  $vars(E)$.  For an expression $E$ and a substitution $\theta$,
  $E\theta$ denotes the \first{instance} of $E$ under $\theta$. An
  expression $E'$ is an instance of another expression $E$ if
  $E'\syneq E\theta$ for some substitution $\theta$ where $A \equiv B$
  denotes that $A$ is syntactically identical to $B$. 
  Let $\theta$ be a substitution and
  $\vcal\subseteq\allvars$.  $\theta\uparrow\vcal\definedas\{X\values
  t\in\theta~|~X\in \vcal\}$ is the \first{restriction} of $\theta$ to
  $\vcal$. The convention is that $\circ$ binds stronger than
  $\uparrow$.  For instance, $\eta\circ\sigma\uparrow V$ is equal to
  $(\eta\circ\sigma)\uparrow V$.

Two substitutions $\sigma$ and $\theta$ are \first{equivalent modulo
  renaming} if there are two renamings $\delta$ and $\rho$ such that
$\sigma=\theta\circ\delta$ and $\theta=\sigma\circ\rho$. We write
$\sigma\cong\theta$ to denote that $\sigma$ and $\theta$ are
equivalent modulo renaming.  We will not distinguish those
substitutions that are equivalent modulo renaming.  $\cong$ is
naturally extended to expressions. Let $E_{1}$ and $E_{2}$ be two
expressions.  $E_{1}\cong E_{2}$ if there are two renamings $\delta$
and $\rho$ such that $E_{1}=E_{2}\delta$ and $E_{2}=E_{1}\rho$.

An \first{equation} is a formula of the form $l=r$ where $l$ and $r$
are terms or atoms. The set of equations is denoted as $Eq$.  Let
$E\in\wp(Eq)$.  $E$ is in \first{solved form} if, for each equation
$l=r$ in $E$, $l$ is a variable and $l$ does not occur in the right
hand side of any equation in $E$. There is a natural bijection between
substitutions and the sets of equations in solved form.  Therefore, we
sometimes write a substitution as a set of equations in solved form.
The \first{unification} of a set of equations is decidable and the
\first{most general unifiers} for a set of equations are equivalent
modulo renaming. Let $mgu$ be the function from $\wp(Eq)$ to
$\{\fail\} \cup (\allvars\mapsto \term(\Sigma,\allvars))$ that, given
a set of equations $E$, either returns a most general unifier for $E$
if $E$ is \first{unifiable} or returns $\fail$ otherwise.
$mgu(\{l=r\})$ is sometimes written as $mgu(l,r)$.

A \first{normal clause} is a formula of the form $H\leftarrow
\Vect{L}$ where $H$ is an atom, $L_{i}$ for each $1\leq i\leq n$ is a
literal. $H$ is called the \first{head} of the clause and $\Vect{L}$
the \first{body} of the clause.  A \first{normal goal} is a formula of
the form $\leftarrow \Vect{L}$ with $L_{i}$ for each $1\leq i\leq n$
being a literal.  A \first{normal program} is a set
$\{C_{\imath}~|~\imath\in \aleph_{C}\}$ of normal clauses where
$\aleph_{C}$ is a finite set of distinct natural numbers.  Let
$m[\imath]$ denote the number of the literals in the body of clause
$C_{\imath}$. We write $C_{\imath}$ as $H_{\imath}\leftarrow
L_{(\imath,1)},L_{(\imath,2)},\cdots,L_{(\imath,m[\imath])}$.

A \first{query} to a program is a goal that initiates the execution of
that program. There might be infinite number of possible queries that
a program is intended to respond to. For the time being, we denote the
set of all possible queries as $\{G_{k}\Theta_{k}~|~k\in\aleph_{G}\}$
where $\aleph_{G}$ is a finite set of distinct natural numbers such
that $\aleph_{G}\cap \aleph_{C}=\emptyset$. $G_{k}$ for each
$k\in\aleph_{G}$ is a normal goal and $\Theta_{k}$ is a set of
substitutions $\theta_{k}$.  Each $G_{k}\theta_{k}$ with
$\theta_{k}\in\Theta_{k}$ is a query. Let $m[k]$ be the number
of literals in $G_{k}$.  We write $G_{k}$ as $\leftarrow
L_{(k,1)},L_{(k,2)},\cdots,L_{(k,m[k])}$.

Let $\aleph\definedas\aleph_{C}\cup\aleph_{G}$. Let $P_{i}$
refer to $C_{i}$ for $i\in\aleph_{C}$ and to refer to $G_{i}$ for
$i\in\aleph_{G}$ and define $\vcal_{i}\definedas vars(P_{i})$.

Let $i\in\aleph$. We designate $P_{i}$ with $m[i]+1$ \first{program
  points}, point $(i,j)$ immediately before $L_{(i,j)}$ for $1\leq
j\leq m[i]$ and point $(i,m[i]+1)$ immediately after $L_{(i,m[i])}$.
$entry(i)\definedas (i,1)$ is called the \first{entry point} of $P_{i}$ and
$exit(i)\definedas (i,m[i]+1)$ the \first{exit point} of $P_{i}$.

We denote by $\points$ the set of all program points designated with
$P_{i}$ for all $i\in\aleph$.  Let $p=(i,j)$ be a program point.
$p[1]=i$ denotes the index to the clause or the query to which $p$
belongs. $p[2]=j$ denotes the position of $p$ in the clause or the query.
So, $p=(p[1],p[2])$. We define two partial functions $\lambda p.p^{+}$
and $\lambda p.p^{\_}$ over the set of all the program
points.\linebreak $p^{+}=(p[1],p[2]+1)$ is defined for each $p$ such
that $p[2]\leq m[p[1]]$ and\linebreak $p^{\_}=(p[1],p[2]-1)$ is
defined for each $p$ satisfying $2\leq p[2]\leq m[p[1]]+1$.  $p^{+}$
is the program point to the right of $p$ if $p^{+}$ exists and
$p^{\_}$ is the program point to the left of $p$ if $p^{\_}$ exists.

Let $p\in \points$. $B_{p}$ denotes the atom in literal $L_{p}$.  If
$L_{p}$ is positive then $L_{p}\equiv B_{p}$. If $L_{p}$ is negative
then $L_{p}\equiv\neg B_{p}$.

\begin{example} {\em  The following normal logic program will be used in
    several examples in this paper. The meaning of $member(X,L)$ is
    that $X$ is a member of list $L$. The meaning of $diff(X,L,K)$ is
    that either $X$ is a member of list $L$ or $X$ is a member of list
    $K$ but $X$ is not both a member of $L$ and a member of $K$.

\[ \begin{array}{lll} 
       C_{1}  &\syneq& diff(X, L, K ) \leftarrow member(X, L),\neg
                       member(X, K)\\
       C_{2}  &\syneq& diff(X, L, K) \leftarrow  member(X, K),\neg
                        member(X, L)\\
       C_{3}  &\syneq& member(X,[X|L]) \leftarrow\\
       C_{4}  &\syneq& member(X,[H|L]) \leftarrow member(X, L)\\
   \end{array}
   \] Suppose that the set of queries is described by
   $\{G_{5}\Theta_{5}\}$ with $G_{5}\syneq \leftarrow diff(X,Y,Z)$ and
   $\Theta_{5}$ is the set of substitutions $\theta$ such that
   $X\theta$ is a variable, and both $L\theta$ and $K\theta$ are
   ground terms. Then, $\aleph_{C}=\{1,2,3,4\}$ and
   $\aleph_{G}=\{5\}$. $\points$ contains 11 program points. $L_{(1,1)}
   = member(X, L)$ and $L_{(1,2)} = \neg member(X, K)$.
   $\vcal_{1}=\{X,L,K\}$ and $\vcal_{5}=\{X,Y,Z\}$.
~\hspace*{\fill}
  $\rule{1ex}{1.5ex}$\par  }
 \label{ex:program}
\end{example} 

\section {Operational semantics} \label{scheme:section:operational}

The operational semantics we consider is the SLDNF-resolution via the
left to right computation rule.  We shall not mention the computation
rule explicitly.  When the set of all the descendant goals of a set of
queries is to be computed, it is necessary to resolve the current goal
with every clause whose head unifies with the selected atom in the
current goal. Therefore, the selection rule for choosing a particular
clause to resolve with the current goal is not of interest.
\cite{Cousot:JLP92} uses general SLD-resolution as operational
semantics of definite logic programs and does not take the computation
rule into account. This usually leads to less precise
analyses~\cite{Janssens:JLP92}.

\subsection{SLDNF} 

We now briefly recall on SLDNF-resolution (SLDNF in short). The
renamed literals will be written in the form $L\rho$ where $L$ is a
literal in a clause or a query and $\rho$ a renaming used in
standardisation apart~\cite{Apt:90}.

First consider \first{SLD-resolution} (SLD in short) for definite
logic programs where every literal is
positive. For a query and a definite program, SLD works by repeatedly
\first{resolving} the current goal, initially the query, with a
clause in the program. In one
resolution step, SLD nondeterministically selects a clause in the
program, renames the clause so that it does not have any common
variable with the current goal, and derives a new goal by replacing
the leftmost literal in the current goal with the body of the renamed
clause and applying to the resultant the most general unifier of the
head of the renamed clause and the leftmost literal. 

If SLD has derived an empty goal, it has successfully computed a
\first{computed answer substitution} to the query. The computed answer
substitution is the restriction, to the variables in the query, of the
composition of the most general unifiers in the derivation from the
query to the empty goal.

\comments{Although there are cases where SLD does not stop either with
  a computed answer substitution or with an indication of failure, we
  are not concerned with the problem of non-termination for the time
  being.}

With SLD, only positive information can be derived from a program. 
  SLDNF uses the \first{negation as failure} rule to
derive negative information.  SLDNF deals with positive literals in
the same way as SLD.
Suppose that the leftmost literal in the current goal is negative.
SLDNF first recursively invokes itself with the leftmost literal as the
query of the recursive invocation.  If the recursive invocation fails
then SLDNF removes the leftmost literal from the current goal and
continues with the resultant as the new current goal.  Otherwise,
according to the rule for the \first{weak safe uses of negation as
  failure}~\cite{Lloyd:87}, SLDNF either backtracks if the computed
answer substitution returned by the recursive invocation is a renaming
or otherwise \first{flounders}.

\comments{The use of negation as failure in a normal program must be safe or
weakly safe~\cite{Lloyd:87a}.  The safe use of negation as failure
demands that if the leftmost literal in the current goal is negative
then it is ground. The weakly safe use of negation as failure demands
that if the leftmost literal in the current goal is negative then the
recursive invocation either fails or returns a renaming as the
computed answer substitution to the negative literal.}

Let the current goal be\\

\(\leftarrow (L_{(i_j,k)}\rho_{j},L_{(i_j,k+1)}\rho_{j},\cdots,
L_{(i_{j},m[i_{j}])}\rho_{j},\cdots)\tau_{(j,k)}\) \hspace{\fill}
(R1)\label{R1}\\ 

\parindent 0pc

where $\rho_{j}$ is the renaming used to rename clause $C_{i_j}$.  If
$L_{(i_j,k)}\syneq\neg B_{(i_j,k)}$ then SLDNF recursively invokes
itself with \( \leftarrow B_{(i_j,k)}\tau_{(j,k)}\) as the query of
the recursive invocation.

\parindent 1pc

Let \mbox{$L_{(i_j,k)}\syneq B_{(i_j,k)}$}.  If
there is a clause \(C_{i_{j+1}}\syneq H_{i_{j+1}}\leftarrow
L_{(i_{j+1},1)},L_{(i_{j+1},2)},\cdots,L_{(i_{j+1},m[i_{j+1}])}\) in
the program and a renaming $\rho_{{j+1}}$ such that
\begin{equation}\label{loc:A}
  vars(C_{i_{j+1}}\rho_{{j+1}})\cap vars(
  (L_{(i_j,k)}\rho_{j},L_{(i_j,k+1)}\rho_{j},\cdots)\tau_{(j,k)})
  =\emptyset
\end{equation}
and $B_{(i_j,k)}\rho_{j}\tau_{(j,k)}$ and $H_{i_{j+1}}\rho_{{j+1}}$
unify, then the new current goal becomes\\ 

$\leftarrow (L_{(i_{j+1},1)}\rho_{{j+1}},\cdots,
L_{(i_{j+1},m[i_{j+1}])}\rho_{{j+1}}, L_{(i_j,k+1)}\rho_{j},\cdots)
\tau_{({j+1},1)}$ \hspace{\fill} (R2)\label{R2}\\ 

\parindent 0pc

where
\begin{equation}\label{loc:B}
  \tau_{({j+1},1)}=\tau_{(j,k)}\circ \eta
\end{equation}
and
\begin{equation}\label{loc:C}
  \eta = mgu(H_{i_{j+1}}\rho_{{j+1}},B_{(i_j,k)}\rho_{j}\tau_{(j,k)})
\end{equation}

\parindent 1pc

Suppose that there is a sub-refutation of \[\leftarrow
(L_{(i_{j+1},1)}\rho_{{j+1}},L_{(i_{j+1},2)}\rho_{{j+1}},\cdots\ 
,L_{(i_{j+1},m[i_{j+1}])}\rho_{{j+1}})\tau_{({j+1},1)}\] with the
composition of the most general unifiers used in the sub-refutation
being $\theta$. Then the next current goal immediately after the
sub-refutation is\\ 

$\leftarrow (L_{(i_j,k+1)}\rho_{j},L_{(i_j,k+2)}\rho_{j},
\cdots,L_{(i_{j},m[i_{j}])}\rho_{j},\cdots)\tau_{(j,k+1)}$
\hspace{\fill} (R3)\label{R3}\\ 

\parindent 0pc

where
\begin{equation}\label{loc:D}
  \tau_{(j,k+1)} = \tau_{({j+1},1)}\circ\theta = \tau_{(j,k)}\circ
  \eta \circ\theta \end{equation} \parindent 1pc

\subsection {VSLDNF}

We now propose a variant of SLDNF (VSLDNF in abbreviation) as the
operational semantics.  VSLDNF is equivalent to SLDNF in the sense
that, given the same goal and the same program, VSLDNF reaches a
program point iff SLDNF reaches the same program point, and the
instantiation of the variables in the clause of the program point by
VSLDNF is equivalent (modulo renaming) to that by SLDNF.  We now
formulate VSLDNF and then establish the equivalence between SLDNF and
VSLDNF.

Let the current goal be\\ 

$\leftarrow
(L_{(i_j,k)},L_{(i_j,k+1)},\cdots,L_{(i_j,m[i_j])})\sigma_{(j,k)}$
\hspace{\fill} (R1')\label{R11}\\

\parindent 0pc

If $L_{(i_j,k)}\equiv\neg B_{(i_j,k)}$ then VSLDNF recursively invokes
itself with \( \leftarrow B_{(i_j,k)}\sigma_{(j,k)} \) being the query
of the recursive invocation.

\parindent 1pc

Let $L_{(i_j,k)}\syneq B_{(i_j,k)}$.  The derivation is suspended and
a sub-derivation is started as follows.  If there is a clause
\(C_{i_{j+1}}\equiv H_{i_{j+1}}\leftarrow
L_{(i_{j+1},1)},L_{(i_{j+1},2)},\cdots,L_{(i_{j+1},m[i_{j+1}])}\) in
the program and a renaming $\psi_{j+1}$ such that
\begin{equation}\label{loc:E} 
\vcal_{i_{j+1}}\cap vars(L_{(i_{j},k)}\sigma_{(j,k)}\psi_{j+1})=\emptyset
\end{equation} and
$B_{(i_{j},k)}\sigma_{(j,k)}\psi_{j+1}$ and $H_{i_{j+1}}$ unify then
the next  current goal becomes\\ 

$(\leftarrow
L_{(i_{j+1},1)},L_{(i_{j+1},2)},\cdots,
L_{(i_{j+1},m[i_{j+1}])})\sigma_{({j+1},1)}$ \hspace{\fill} (R2')\label{R21}\\ 

\parindent 0pc

where
\begin{equation}\label{loc:F}
\sigma_{(j+1,1)} =
mgu(B_{(i_{j},k)}\sigma_{(j,k)}\psi_{j+1},H_{i_{j+1}}) 
\end{equation}
We call the step to derive a new goal from the current goal and a
clause in the program \first{procedure-entry}.

\parindent 1pc

Suppose that there is a sub-refutation of
\[(\leftarrow
L_{(i_{j+1},1)},L_{(i_{j+1},2)},\cdots,L_{(i_{j+1},m[i_{j+1}])})\sigma_{(j+1,1)}\]
and $\sigma_{(j+1,m[i_{j+1}]+1)}$ be the substitution immediately
after the sub-refutation. Then the suspended derivation is resumed and
the new current goal becomes \\

$\leftarrow
  (L_{(i_{j},k+1)},\cdots,L_{(i_{j},m[i_{j}])})\sigma_{(j,k+1)}$
  \hspace{\fill} (R3')\label{R31}\\

\parindent 0pc

where
\begin{equation}\label{loc:G}
  \sigma_{(j,k+1)}=\sigma_{(j,k)}\circ
  mgu(B_{(i_{j},k)}\sigma_{(j,k)},H_{i_{j+1}}\sigma_{({j+1},m[i_{j+1}]+1)}\phi_{j+1})
\end{equation} 
and $\phi_{j+1}$ is a renaming such that 
\begin{equation}\label{loc:H}
vars(H_{i_{j+1}}\sigma_{(j+1,m[i_{j+1}]+1)}\phi_{j+1})\cap
vars(C_{i_{j}}\sigma_{{j,k}})=\emptyset
\end{equation}
We call the step to derive a new goal from a suspended goal and a
completed sub-derivation \first{procedure-exit}. 

\parindent 1pc

\begin{lemma}  \label{lm:scheme:vsldnf}
   {\em
    VSLDNF is equivalent to SLDNF in the sense that, given the same
    goal and the same program, VSLDNF reaches a program point iff
    SLDNF reaches the same program point, and the instantiation of the
    variables in the clause of the program point by VSLDNF is
    equivalent (modulo renaming) to that by SLDNF.}

\begin{proof} See p.\pageref{pr:lm:scheme:vsldnf}.\end{proof} 
\end{lemma} 

\begin{example} {\em This example illustrates VSLDNF. 
    Let the program be that in example~\ref{ex:program} and
    $\leftarrow diff(X,[2,1],[3,1])$ be the query. Let
    $\sigma_{(0,1)}=\emptyset$. VSLDNF begins with the following
    current goal.\\ 

\( \leftarrow diff(X,[2,1],[3,1]) \sigma_{(0,1)}\)\hspace{\fill} (G0)\\

Let $\psi_{1}=\{X\values X_{1}\}$ and $C_{i_{1}}=C_{1}$. Then 
\[\begin{array}{lll} 
     \sigma_{(1,1)} &=& mgu(diff(X,[2,1],[3,1])
     \sigma_{(0,1)}\psi_{1},diff(X,L,K))\\
     &=& \{X_{1}\values X, L\values [2,1],K\values [3,1]\} \\
  \end{array}
\] VSLDNF suspends goal (G0), performs a
procedure-entry, and derives the
following goal.\\

\( \leftarrow (member(X,L),\neg member(X,K))\sigma_{(1,1)}\)
\hspace{\fill} (G1)\\

Let $\psi_{2}=\{X\values X_{2}\}$ and $C_{i_{2}}=C_{3}$. Then 
\[\begin{array}{lll}
      \sigma_{(2,1)} &=&
      mgu(member(X,L)\sigma_{(1,1)}\psi_{2},member(X,[X|L]))\\
      &=& \{X_{2}\values 2, X\values 2, L\values [1]\}
  \end{array}
  \] VSLDNF suspends goal (G1), performs a
  procedure-entry, and derives the following empty goal.\\ 

  \( \Box \sigma_{(2,1)}\)\hspace{\fill} (G2)\\

  VSLDNF performs a procedure-exit step and derives the following from
  (G1).\\ 

  \( \leftarrow \neg member(X,K))\sigma_{(1,2)}\)
\hspace{\fill} (G3)\\ 

\parindent 0pc

where, letting $\phi_{2}=\emptyset$,
\[\begin{array}{lll}
      \sigma_{(1,2)}  &=& \sigma_{(1,1)}\circ
      mgu(member(X,L)\sigma_{(1,1)},
      member(X,[X|L])\sigma_{(2,1)}\phi_{2})\\ 
      &=& \{X_1\values 2, X\values 2,L\values [2,1], K\values [3,1]\}
  \end{array}
\] 

\parindent 1pc

The leftmost literal of (G3) is negative, VSLDNF invokes itself
recursively with\linebreak $\leftarrow member(X,K))\sigma_{(1,2)}$ and fails to
refute it. So, by the negation as failure rule, VSLDNF derives the
following goal with $\sigma_{(1,3)}=\sigma_{(1,2)}$.\\ 

\( \Box \sigma_{(1,3)}\)\hspace{\fill} (G4)\\ 

This finishes a sub-refutation of (G1). VSLDNF performs a
procedure-exit step and derives the following from (G0).

  \( \Box  \sigma_{(0,2)}\)\hspace{\fill} (G5)\\

\parindent 0pc

where, letting $\phi_{1}=\emptyset$,

\[\begin{array}{lll} \sigma_{(0,2)} &=& \sigma_{(0,1)}\circ
mgu(diff(X,[2,1],[3,1])\sigma_{(0,1)},
diff(X,L,K)\sigma_{(1,3)}\phi_{1})\\ &=& \{X\values 2\}
  \end{array}
  \]

\parindent 1pc

This finishes a refutation of (G0). So, $\sigma_{(0,2)} = \{X\values
2\}$ is a computed answer substitution of $\leftarrow
diff(X,[2,1],[3,1])$. $\Box$}
\end{example}

VSLDNF differs from SLDNF in several ways.  Firstly, a goal in VSLDNF
is a part of a clause or a query, in particular, it is a tail of a
clause or a query. This helps to approximate VSLDNF as a transition
system later.  Secondly, when VSLDNF derives a new goal from the
current goal and a clause, it renames the leftmost literal in the goal
instead of the clause.  This is to ensure that the domain of the
substitution that will be applied to the body of the clause contains
variables in the clause instead of their renamed counterparts.
Thirdly, when a sub-refutation is finished, an extra renaming and an
extra unification are needed for VSLDNF to calculate the substitution
immediately after the sub-refutation whilst these extra operations are
not needed in SLDNF. Note that VSLDNF is only used in formulating the
collecting semantics.

\subsection {Program graph} 
  
Let $p,q\in\points$, and $q$ be the most recent program point that
VSLDNF has reached. There are several possibilities that  VSLDNF
will reach $p$ next. If $q$ is the exit point of a clause then the
only way that VSLDNF can reach $p$ immediately is to perform a
procedure-exit.  This can happen only if $L_{p^{\_}}$ is positive and
that program clause has been used to resolve with $L_{p^{\_}}$. If $q$ is not the exit point of a clause then VSLDNF may reach $p$ immediately either by performing a procedure-entry or by applying the negation as failure rule.  VSLDNF may reach $p$ immediately by applying negation as failure rule if $L_{q}\equiv \neg B_{q} \wedge q=p^{\_}$.  VSLDNF may reach $p$ immediately by performing a procedure-entry either directly when $L_{q}$ is positive or indirectly when $L_{q}$ is negative.  Note that if $q$ is the exit point of a query then VSLDNF has succeeded and will not visit any more program points. In order to facilitate further presentation, we assume that VSLDNF starts at a dummy program point $(0,0)\not\in \points$ from where it can reach entry points of goal clauses by doing nothing.  Therefore, there are four ways that VSLDNF will reach $p$ immediately after it has reached $q$.  We use a graph $<\points^{+},\edges>$, called program graph, to represent the relation among program points $p,q$ that ``VSLDNF will possibly visit $p$ immediately after it has visited $q$.  The set $\points^{+}$ of nodes in the program graph is $\points\cup\{(0,0)\}$ and each edge $\edge{p}{q}$ in $\edges$ in the program graph denotes that VSLDNF will possibly visit $p$ immediately after it has visited $q$.  Formally, $\edges$ is inductively defined as follows.  \begin{eqnarray*}
  \edges &\definedas& \bigcup_{0\leq\jmath\leq 3} \edges^{\jmath}\\ \edges^{0} &\definedas&
  \{\edge{entry(k)}{(0,0)}~|~k\in\aleph_{G}\}\\ \edges^{1} &\definedas&
  \left\{\edge{entry(i)}{q} ~\begin{array}{|ll} & q[2]\leq m[q[1]]\\ \wedge &
  i\in\aleph_{C}\\ \wedge & \exists \rho. \left( \begin{array}{ll} &
  \rho~is~a~renaming\\ \wedge & vars(B_{q}\rho)\cap
  vars(H_{i})=\emptyset\\ \wedge & mgu(B_{q}\rho,H_{i})\neq\fail
                  \end{array} \right)
                                           \end{array}
                                         \right\}\\ \edges^{2}
                                         &\definedas& \{
                                         \edge{p}{exit(i)}~|~\edge{entry(i)}{p^{\_}}>
                                         \in \edges^{1}\wedge
                                         L_{p^{\_}} \equiv B_{p^{\_}} \}\\ 
                                         \edges^{3} &\definedas&
                                         \{\edge{p}{p^{\_}}~|~L_{p^{\_}}\equiv \neg
                                         B_{p^{\_}}\}\\ 
\end{eqnarray*}

$\edges^{0}$ is the collection of edges from dummy program point
$(0,0)$ to the entry points of queries. Edges in $\edges^{1}$
correspond to procedure-entries, edges in $\edges^{2}$ to
procedure-exits, and edges in $\edges^{3}$ to the negation as failure
rule.  $\edges^{\imath}\cap \edges^{\jmath}=\emptyset$ for any
$0\leq\imath,\jmath\leq 3$ such that $\imath\neq \jmath$.

\begin{example} {\em Consider the program in example~\ref{ex:program}.

    $5\in\aleph_{G}$. Hence, $\edge{(5,1)}{(0,0)}\in \edges$.  Let
    $\rho=\{X\values X_{0},L\values L_{0}\}$.\linebreak
    $mgu(B_{(1,1)}\rho,H_{3})=\{X_{0}\values X,L_{0}\values
    [X|L]\}\neq\fail$. So, \mbox{$\edge{(3,1)}{(1,1)}\in\edges$}.
    Since $L_{(1,1)}$ is positive,
    \mbox{$\edge{(1,2)}{(3,1)}\in\edges$}.  Let
    \mbox{$\delta=\{X\values X_{0},K\values K_{0}\}$}.
    \mbox{$mgu(B_{(1,2)}\delta,H_{3})=\{X_{0}\values X,K_{0}\values
      [X|L]\}\neq\fail$}. So, $\edge{(3,1)}{(1,2)}\in\edges$.  Since
    $L_{(1,2)}$ is negative, $\edge{(1,3)}{(1,2)}\in\edges$.
 
    There are 23 edges in the program graph for the program. $\Box$}
\end{example}

\section {Collecting semantics} \label{scheme:section:static}

In this section, we present the fixed-point collecting semantics for
normal programs. The collecting semantics of normal program $P$ is
$\lfp \ssem$ where $\ssem$ is defined
below. $\lfp \ssem$ associates a set of substitutions
with each edge $\edge{p}{q}\in \edges$.
Sub-section~\ref{subsec:transition} uses a transition system to
approximate VSLDNF. A state in this transition system corresponds to a
goal in VSLDNF. The set of states derivable from a set of initial
states by the transition system is then characterised as the least
fixed-point $\lfp \sem$ of a function $\sem$ mapping
a set of states into another set of states. Therefore, the set of
goals derivable from a set of initial goals by VSLDNF is approximated
by $\lfp \sem$.  Sub-section~\ref{subsec:base} derives
the fixed-point collecting semantic function $\ssem$ from
$\sem$ and proves that $\lfp \ssem$ is a safe
approximation of $\lfp \sem$. 

Let $A,B$ be atoms, and $\theta,\omega\in Sub$. Define
\begin{equation} 
\unify(A,\theta,B,\omega)\definedas
  \left\{\begin{array} {l}
    let~\rho~be~a~renaming~such~that~vars(A\theta\rho)\cap
    vars(B\omega)=\emptyset,\\
    if~~~~mgu(A\theta\rho,B\omega)\neq \fail\\
    then~\omega\circ mgu(A\theta\rho,B\omega)\\
    else~~\fail
     \end{array} \right.
   \label{fr:eq:unify}
\end{equation}

Although there are infinite number of renamings $\rho$ satisfying
$vars(A\theta\rho)\cap vars(B\omega)=\emptyset$ in
equation~\ref{fr:eq:unify}, only one renaming must be considered when
computing $\unify(A,\theta,B,\omega)$ because
$\unify(A,\theta,B,\omega)$ for one renaming is equivalent (modulo
renaming) to $\unify(A,\theta,B,\omega)$ for another renaming accoding
to lemma~\ref{appendix:lemma:renaming}.

\subsection{Approximating VSLDNF by a transition  system}
\label{subsec:transition} 

We now devise a transition system to approximate VSLDNF.  A state in
the transition system corresponds to a goal in VSLDNF. The transition
system approximates VSLDNF in the sense that, if a goal is derivable
from an initial goal by VSLDNF then the state corresponding to the
goal is derivable from the state corresponding to the initial goal by
the transition system while the reverse is not necessarily true.

A state in the transition system is a stack that is a sequence of
stack items.  The empty stack is denoted as $\$$.  A stack item is of
the form \stackitem{p}{q}{\theta} where $\edge{p}{q}\in\edges$ and
$\theta\in Sub$.  The meaning of \stackitem{p}{q}{\theta} is that the
control of execution transfers from $q$ to $p$ with $\theta$ being the
substitution at $p$. The set $\stackitems$ of all  possible
stack items is therefore
\[ \stackitems = \{\stackitem{p}{q}{\theta}~|~ \edge{p}{q}\in \edges
\wedge \theta\in Sub\}
\] 

The set $\stacks$ of all possible stacks is the set of all possible
sequences of stack items from $\stackitems$. $\stacks$ can be
inductively defined as follows.
\begin{itemize}
\item $\$\in\stacks$; and 
\item $\stackitem{p}{q}{\theta}\cdot S \in\stacks$ if $\stackitem{p}{q}{\theta}\in
  \stackitems \wedge S \in \stacks$.
\end{itemize}

Let \mbox{$x_{1}\in\stackitems,\ldots,x_{n}\in \stackitems$} and
$S\in\stacks$.  \mbox{$x_{1}\cdot\ldots\cdot x_{n}\cdot{S}$} is
sometimes written as
\[ \begin{array} {c}
    \underline{x_{1}}\\ \vdots \\
    \overline{\underline{x_{n}}}\\ S
   \end{array} 
\]

The set $\initstacks\subseteq \stacks$ of initial states is
determined by the set of queries in VSLDNF.
\[  \initstacks\definedas
\{\stackitem{p}{q}{\theta}\cdot\$~|~\stackitem{p}{q}{\theta}\in\stackitems\wedge 
  \edge{p}{q}\in \edges^{0}\wedge \theta\in\Theta_{p[1]}\}
\]

The set of final states is 
\[ \finalstacks = \{\stackitem{exit(k)}{q}{\theta}\cdot\$~|~
k\in\aleph_{G}\wedge \theta \in Sub \wedge \edge{exit(k)}{q}\in
\edges\}
\]

The set of  descendant states of the set $\initstacks$ of initial states
is obtained by applying the transition rules in
figure~\ref{fig:rules}.  Rule (0) says that every final state is
stable.  Rule (1a) corresponds to direct procedure-entry and rule (1b)
to indirect procedure-entry. Rule (2) corresponds to procedure-exit
and rule (3) deals with negative literals.

Rule (3) causes the inaccuracy of the transition system. When the
transition system reaches a state corresponding to a goal with its
leftmost literal being negative, the transition system may apply
either rule (1b) or rule (3). Applying rule (1b), it will go to a
state corresponding to a goal after performing a procedure-entry
indirectly.  The application of rule (1b) is to enable information to
propagate forward so as to ensure that the transition system safely
approximates VSLDNF. Applying rule (3), it will go to a state
corresponding to the goal as if the recursive invocation of VSLDNF
with the negative literal had failed while the recursive invocation
may succeed in some cases. This results in a simple approximation of
VSLDNF.

\begin{figure}
\begin{center}
\begin{tabular}{||c|c||} \hline\hline
Rule (0) & $\begin{array}{ll}
  If &
    k\in\aleph_{G} \\
  then &
   \stackitem{exit(k)}{q}{\theta}\cdot\$ \transfers
   \stackitem{exit(k)}{q}{\theta}\cdot\$
 \end{array} $\\ \hline
Rule (1a) & $\begin{array}{ll} 
  If &  \left(\begin{array}{ll}
            &  L_{q} \equiv B_{q} \wedge \edge{entry(i)}{q}\in \edges^{1}\\
           \wedge & \theta=\unify(B_{q},\sigma,H_{i},\epsilon)\neq \fail
        \end{array}\right) \\
  then & \stackitem{q}{u}{\sigma}\cdot{S} \transfers
         \stackitem{entry(i)}{q}{\theta}\cdot\stackitem{q}{u}{\sigma}\cdot{S}
  \end{array} $ \\ \hline
Rule (1b) & $\begin{array}{ll} 
     If & \left(\begin{array}{ll}
            &  L_{q} \equiv \neg B_{q} \wedge \edge{entry(i)}{q}\in \edges^{1}\\
           \wedge & \theta=\unify(B_{q},\sigma,H_{i},\epsilon)\neq \fail
          \end{array}\right) \\
      then & ~\stackitem{q}{u}{\sigma}\cdot{S} \transfers \stackitem{entry(i)}{q}{\theta}\cdot\$
     \end{array}$\\ \hline
Rule (2) & $\begin{array}{ll}
  If & \left(
   \begin{array}{ll}
    & \edge{p}{q}\in \edges^{2}\\
    \wedge & \theta= \unify(H_{q[1]},\sigma,B_{p^{\_}},\eta)\neq \fail
   \end{array}\right)\\
  then &
   ~\stackitem{q}{u}{\sigma}\cdot\stackitem{p^{\_}}{v}{\eta}\cdot{S}
   \transfers \stackitem{p}{q}{\theta}\cdot{S}
 \end{array} $\\\hline
Rule (3) & $\begin{array}{ll}
  If &
    \edge{p}{q}\in \edges^{3}\\
  then &
   \stackitem{q}{u}{\theta}\cdot{S}
   \transfers \stackitem{p}{q}{\theta}\cdot{S}
 \end{array}$ \\\hline\hline
\end{tabular} \end{center}
 \caption{\label{fig:rules} Transition rules}
\end{figure}

The set of  descendant states of a set $\initstacks$ of initial states is
therefore the least fixed-point of function $\sem$ that is defined as
follows.

\begin{eqnarray} \label{eq:tsem} 
  \sem(X) & \definedas & \bigcup_{0\leq\jmath\leq 3} \sem^{\jmath}(X)\\
 \sem^{0}(X) & \definedas & \{ {\stackitem{p}{q}{\theta}}\cdot\$~|~
           \edge{p}{q}\in \edges^{0} 
          \wedge  \theta\in \Theta_{p[1]}
         \} \label{eq:tsem:2}\\
  \sem^{1}(X) & \definedas & \label{eq:tsem:1} 
      \left\{
           \begin{array}{c} \underline{\stackitem{p}{q}{\theta}}\\
             \underline{\stackitem{q}{u}{\sigma}}\\ {S}
           \end{array}
       \begin{array}{|ll}
         & \edge{p}{q}\in \edges^{1}
         \wedge  L_{q}\syneq B_{q} \\
         \wedge & \stackitem{q}{u}{\sigma}\cdot{S}\in X\\ 
         \wedge &
         \theta=\unify(B_{q},\sigma,H_{p[1]},\epsilon)\neq \fail
       \end{array} 
     \right\}\\ 
    & \cup & \left\{ {\stackitem{p}{q}{\theta}}\cdot\$
       \begin{array}{|ll}
         & \edge{p}{q}\in \edges^{1}
         \wedge  L_{q} \syneq \neg B_{q} \\
         \wedge & \stackitem{q}{u}{\sigma}\cdot{S}\in X\\ 
         \wedge &
         \theta=\unify(B_{q},\sigma,H_{p[1]},\epsilon)\neq \fail
       \end{array} 
     \right\}\nonumber\\ 
 \sem^{2}(X) & \definedas & \label{eq:tsem:3}
    \left\{ \stackitem{p}{q}{\theta}\cdot{S}
\begin{array}{|ll} 
   & \edge{p}{q}\in \edges^{2}\\
   \wedge & \begin{array}{c} \underline{\stackitem{q}{u}{\sigma}}\\
                             \underline{\stackitem{p^{\_}}{v}{\eta}}\\ 
                            {S}
            \end{array} \in  X\\
  \wedge & \theta=\unify(H_{q[1]},\sigma,B_{p^{\_}},\eta)\neq\fail
\end{array}
\right\}\\ 
\sem^{3}(X) &\definedas& \{ \stackitem{p}{q}{\theta}\cdot S~|~
     \edge{p}{q}\in \edges^{3}
    \wedge   \stackitem{q}{u}{\theta}\cdot S\in X 
  \}
  \label{eq:tsem:4}
\end{eqnarray}

The domain $\dom$ of $\sem$ is $\wp(\stacks)$. $<\dom,
\subseteq>$ is a complete lattice and
$\sem$ is monotonic on
$<\dom,\subseteq>$.

Rule (0) of the transition system is not embodied in $\sem$.
Since $\sem$ is a monotonic function on $<\dom,\subseteq>$, we
have $\sem\uparrow k\subseteq \sem\uparrow (k+1)$ for any
$k\geq 0$. Therefore, any final state will be in $\lfp
\sem$ if it is derivable from an initial state by the transition
system.

\subsection {Collecting semantics} 
\label{subsec:base}

$\lfp \sem$ is a set of states. A state is a stack that
corresponds to a goal.  The collecting semantics $\lfp
\ssem$ first abstracts away the sequential relation
between stack items of a stack and then classifies the
stack items according to edges $\edge{p}{q}$. $\lfp
\ssem$  associates each
$\edge{p}{q}$ in $\edges$ with the set of substitutions $\theta$ such
 that \mbox{$\stackitem{p}{q}{\theta}$} is a stack item in
$\stackitems$. 
Each edge $\edge{p}{q}\in\edges$ will be assocaited
with a member from $\wp(Sub)$.  $<\wp(Sub),\subseteq,\emptyset,Sub,
\cap,\cup>$ is a complete lattics.  Therefore, the domain
$\sdom$ of the collecting semantics is the Cartesian product of
the same component domain $\wp(Sub)$ for as many times as the number
of edges $\edge{p}{q}$ in $\edges$.  Let $\sbigx\in\sdom$. We
use \mbox{$\sbigx_{\edge{p}{q}}$} to denote the component in
$\sbigx$ that corresponds to edge $\edge{p}{q}$.  Let
$\sbigx,\sbigy\in\sdom$. Define
\[\begin {array} {lll}  
\sbigx \ssqsubseteq \sbigy & \definedas &
    \forall\edge{p}{q}\in \edges.
                    (\sbigx_{\edge{p}{q}} \subseteq 
                    \sbigy_{\edge{p}{q}})\\
{[\sbigx \ssqcap \sbigy]}_{\edge{p}{q}} &
    \definedas & \sbigx_{\edge{p}{q}}\cap \sbigy_{\edge{p}{q}} \\
{[\sbigx \ssqcup \sbigy]}_{\edge{p}{q}} &
    \definedas & \sbigx_{\edge{p}{q}}\cup \sbigy_{\edge{p}{q}} \\
\stop_{\edge{p}{q}} &\definedas & Sub\\
\sbot_{\edge{p}{q}} &\definedas & \emptyset
\end{array}
\] $<\sdom,\ssqsubseteq, \sbot, \stop,
\ssqcap, \ssqcup>$ is a complete lattice.

The approximation of a set of stacks by a vector of sets of
substitutions is modeled by the following monotonic function
$\sgamma\in \sdom\mapsto\dom$.

\begin{equation}\label{gammaint}
  \sgamma(\sbigx) = \left\{
   \begin{array}{c|}
  \underline{\stackitem{p_{1}}{q_{1}}{\theta_{1}}}\\
  \vdots\\
  \underline{\stackitem{p_{n}}{q_{n}}{\theta_{n}}}\\
  \$
  \end{array} 
  ~~\forall 1\leq i\leq n.
  (\edge{p_{i}}{q_{i}}\in\edges\wedge \theta_{i}\in
  \sbigx_{\edge{p_{i}}{q_{i}}}) \right\} 
\end{equation}

Let $A,B$ be atoms, and $\Theta,\Omega$ be sets of substitutions. Define
\begin{equation}
   \sunify (A,\Theta,B,\Omega) \definedas\{\unify(A,\theta,B,\omega)\neq
  \fail~|~\theta\in\Theta\wedge\omega\in\Omega\}
\end{equation} 
The fixed-point collecting semantics is defined in the following.

\begin{eqnarray} 
  \lefteqn{ {[ \ssem(\sbigx)]}_{\edge{p}{q}} \definedas}
  \nonumber\\ & & \label{eq:csem:2} \Theta_{p[1]}
  ~~~~~~~~~~~~~~~~~~~~~~~~~~~~~~~~~~~~~~~~~~~~~~~~~~~~~~~~~~~~~~~~~~~~~~~~~~~
  if~\edge{p}{q}\in\edges^{0}\\ & & \label{eq:csem:1}
  \bigcup \{ \sunify
  (B_{q},\sbigx_{\edge{q}{u}},H_{p[1]},\{\epsilon\})~|~\edge{q}{u}\in\edges\} ~~~~~~~~~~~~~~~~~~~if~
  \edge{p}{q}\in\edges^{1}\\ & & \label{eq:csem:3}
  \bigcup \left\{
  \sunify(H_{q[1]},\sbigx_{\edge{q}{u}},
  B_{p^{\_}},\sbigx_{\edge{p^{\_}}{v}})~\begin{array}{|ll}
        & {\edge{p^{\_}}{v}\in \edges} \\
        \wedge & {\edge{q}{u}\in\edges}\end{array} \right\}~~~if~
  \edge{p}{q}\in\edges^{2}\\ & & \label{eq:csem:4}
  \bigcup
   \{ \sbigx_{\edge{q}{u}}~|~\edge{q}{u}\in \edges\}~~~~~~~~~~~~~~~~~~~~~~~~~~~~~~~~~~~~~~~~~~~~~~~~~
  if~\edge{p}{q}\in\edges^{3}
\end{eqnarray} 

$\ssem$ is a monotonic function on
$<\sdom,\ssqsubseteq>$.

\begin{example}\label{ex:collecting} {\em  Let $P$ be the program in
    example~\ref{ex:program}. $\ssem$ is a system of 23
    simultaneous recurrence equations. Each equation corresponds to an
    edge in $\edges$. The following four equations are examples of
    equations~\ref{eq:csem:2}-\ref{eq:csem:4} respectively. Let
    $A=member(X,L)$ and $B=member(X,[X|L])$.

\begin{eqnarray*} 
\lefteqn{{[\ssem(\sbigx)]}_{\edge{(3,1)}{(1,1)}} = } \\
 & \sunify(A,\sbigx_{\edge{(1,1)}{(5,1)}},B,
                 \{\epsilon\})\\
\lefteqn{{[\ssem(\sbigx)]}_{\edge{(5,1)}{(0,0)}} = \Theta_{5}}\\
\lefteqn{{[ \ssem(\sbigx)]}_{\edge{(1,2)}{(3,1)}} = }\\
& \begin{array}{ll} 
 &  \sunify(B,\sbigx_{\edge{(3,1)}{(1,1)}},
                 A,\sbigx_{\edge{(1,1)}{(5,1)}}) \\
 \cup &    \sunify(B,\sbigx_{\edge{(3,1)}{(1,2)}},
                 A,\sbigx_{\edge{(1,1)}{(5,1)}}) \\
 \cup &   \sunify(B,\sbigx_{\edge{(3,1)}{(2,1)}},
                 A,\sbigx_{\edge{(1,1)}{(5,1)}}) \\
 \cup &   \sunify(B,\sbigx_{\edge{(3,1)}{(2,2)}},
                 A,\sbigx_{\edge{(1,1)}{(5,1)}}) \\
 \cup &   \sunify(B,\sbigx_{\edge{(3,1)}{(4,1)}},
                 A,\sbigx_{\edge{(1,1)}{(5,1)}}) 
  \end{array}\\
\lefteqn{{[ \ssem(\sbigx)]}_{\edge{(1,3)}{(1,2)}} = 
    \sbigx_{\edge{(1,2)}{(3,1)}}\cup \sbigx_{\edge{(1,2)}{(4,2)}}}
\end{eqnarray*} 
$\Box$}
\end{example} 

\begin{lemma} \label{lm:scheme:Fsharp}
  {\em $\lfp \sem \subseteq \sgamma(\lfp
    \ssem)$. }

\begin{proof} See p.\pageref{pr:lm:scheme:Fsharp}. \end{proof}
\end{lemma}

\section{The Generic Abstract Semantics $\fsem$} \label{scheme:section:abstract}

The collecting semantics $\lfp \ssem$ is a safe approximation of the
operational semantics and can be used as a basis for program analysis
because any safe approximation of this collecting semantics is a safe
approximation of the operational semantics.  ${[\lfp
\ssem]}_{\edge{p}{q}}$ contains all the substitutions whenever the
control of execution transfers from program point $q$ to program point
$p$.  ${[\lfp \ssem]}_{\edge{p}{q}}$ is usually an infinite set of
substitutions and is therefore not computable in finite time. In order
to obtain useful information about the possible substitutions when the
control of execution transfers from program point $q$ to program point
$p$, further approximations are needed.  This section derives the
generic abstract semantics $\fsem$ from $\ssem$. 

\subsection {Abstract domains} 
The collecting semantics $\lfp \ssem$ associates with each
edge $\edge{p}{q}$ a set of substitutions which is a superset of the
set of the substitutions whenever the control of execution transfers
from $q$ to $p$.
When program is analysed by means of abstract interpretation, the set
of substitutions associated with $\edge{p}{q}$ is approximated by an
abstract substitution associated with $\edge{p}{q}$.  For edge
$\edge{p}{q}$, only values of the variables in $\vcal_{p[1]}$ are of
interest and, for edge $\edge{p'}{q'}$, only values of the variables
in $\vcal_{p'[1]}$ are of interest. The abstract substitions for
$\edge{p}{q}$ and $\edge{p'}{q'}$ are from different domains when
$p'[1]\neq p[1]$.  We will simply call a domain for abstract
substitutions an abstract domain.  We find it convenient to
parameterise abstract domains with finite sets of variables instead of
having a single abstract domain for all abstract substitutions
associated with different edges or constructing abstract domains for
different edges in different ways. Let $\asub{\vcal}$ denote the
domain for abstract substitutions for describing values of variables
in $\vcal$.  Then ${[\lfp \ssem]}_{\edge{p}{q}}$ is
represented by a member of $\asub{\vcal_{p[1]}}$.  We require
that, for any finite $\vcal\subseteq \allvars$,
\begin{itemize} \label{scheme:c1c2}
\item [C1:] $<\asub{\vcal},\asuborder{\vcal},\asubbot{\vcal},
  \asubtop{\vcal}, \asubcap{\vcal}, \asubcup{\vcal}>$ is a complete lattice
  where $\asuborder{\vcal}$ is a partial order on $\asub{\vcal}$,
  $\asubbot{\vcal}$ the infimum, $\asubtop{\vcal}$ the supremum,
  $\asubcap{\vcal}$ the greatest lower bound operator and
  $\asubcup{\vcal}$ the least upper bound operator; and
\item [C2:] there is a monotonic function  
  $\asubgamma{\vcal} \in \asub{\vcal} \mapsto
  \wp(Sub)$.
\end{itemize}  

The domain $\fdom$ of $\fsem$ is constructed in the same
manner as the domain $\sdom$ of $\ssem$ was constructed. Each
member $\fbigx$ in $\fdom$ is a vector that is indexed by edges
$\edge{p}{q}$ in $\edges$.  $\fbigx_{\edge{p}{q}}$ is an element from
$\asub{\vcal_{p[1]}}$.  Let $\fbigx\in \fdom$ and
$\fbigy\in \fdom$.  Define

\[\begin {array} {lll} \fbigx \fsqsubseteq \fbigy & \definedas
& \forall\edge{p}{q}\in \edges.  (\fbigx_{\edge{p}{q}}
~~\asuborder{\vcal_{p[1]}}~~
\fbigy_{\edge{p}{q}})\\ {[\fbigx \fsqcap
  \fbigy]}_{\edge{p}{q}} & \definedas & \fbigx_{\edge{p}{q}}
~~\asubcap{\vcal_{p[1]}}~~ \fbigy_{\edge{p}{q}} \\ 
{[\fbigx \fsqcup \fbigy]}_{\edge{p}{q}} & \definedas &
\fbigx_{\edge{p}{q}} ~~\asubcup{\vcal_{p[1]}}~~
\fbigy_{\edge{p}{q}} \\ \ftop_{\edge{p}{q}} &\definedas &
\asubtop{\vcal_{p[1]}}\\ \fbot_{\edge{p}{q}}
&\definedas & \asubbot{\vcal_{p[1]}}
\end{array}
\]

$<\fdom,\fsqsubseteq,\fbot,\ftop,
\fsqcap,\fsqcup>$ is a complete lattice.

The concretisation function $\fgamma$ is defined in terms of
$\asubgamma{\vcal_{i}}$. For every $\edge{p}{q}\in \edges$ and
$\fbigx\in\fdom$,
\begin{equation} 
  \label{e:concretisation}
  {[\fgamma(\fbigx)]}_{\edge{p}{q}} = \asubgamma{\vcal_{p[1]}}({\fbigx}_{\edge{p}{q}})
\end{equation}

The monotonicity of $\fgamma\in \fdom\mapsto \sdom$ follows
immediately from equation~\ref{e:concretisation}
and C2.

\subsection {The Generic Abstract Semantics $\fsem$} 

$\fsem$ is derived from $\ssem$ as follows.  A set $\Theta\in
\wp(Sub)$ of substitutions is replaced by an abstract substitution
$\ftheta$ in $\asub{\vcal}$ where $\vcal$ is a set of variables
whose values are of interest. $\sunify$ applied to two sets of
substitutions described by $\ftheta\in\asub{\ucal}$ and
$\fsigma\in\asub{\vcal}$ respectively is replaced by an operator
$\aunify{\ucal}{\vcal}$ applied to $\ftheta$ and $\fsigma$.
$\cup$ in the definition of ${[ \ssem]}_{\edge{p}{q}}$ is replaced
by $\asubcup{\vcal_{p[1]}}$.  Let $\ftheta_{k}\in
\asub{\vcal_{k}}$ be the least abstract substitution such that
$\Theta_{k}\subseteq \asubgamma{\vcal_{k}}(\ftheta_{k})$ for each
$k\in\aleph_{G}$. Note that $\ftheta_{k}$ instead of $\Theta_{k}$
is given before the program is analysed.  Let
$\asubid{\vcal_{i}}\in \asub{\vcal_{i}}$, called an abstract
identity substitution in~\cite{Bruynooghe91}, be the least
abstract substitution such that $\epsilon\in
\asubgamma{\vcal_{i}}(\asubid{\vcal_{i}})$ for each
$i\in\aleph_{C}$. $\fsem$ is defined as
follows.

\begin{eqnarray}
  \lefteqn{{[ \fsem(\fbigx)]}_{\edge{p}{q}}\definedas}
  \nonumber\\ 
& &  \ftheta_{p[1]}
~~~~~~~~~~~~~~~~~~~~~~~~~~~~~~~~~~~~~~~~~~~~~~~~~~~~~~~~~~~
   if~\edge{p}{q}\in \edges^{0}  \label{eq:asem:2}\\
 & & \label{eq:asem:1}
 \asubcup{\vcal_{p[1]}}\{ \aunify{\vcal_{q[1]}}{\vcal_{p[1]}}(B_{q},\fbigx_{\edge{q}{u}},H_{p[1]},
  \asubid{\vcal_{p[1]}})
  |\edge{q}{u}\in\edges\}~\\
& &~~~~~~~~~~~~~~~~~~~~~~~~~~~~~~~~~~~~~~~~~~~~~~~~~~~~~~~~~~~~~~~~~
   if~ \edge{p}{q}\in \edges^{1}\nonumber\\
& & \label{eq:asem:3}
 \asubcup{\vcal_{p[1]}} \left\{
  \aunify{\vcal_{q[1]}}{\vcal_{p[1]}}(H_{q[1]},\fbigx_{\edge{q}{u}},B_{p^{\_}},\fbigx_{\edge{p^{\_}}{v}})
 \begin{array}{|l}
  ~~~\edge{p^{\_}}{v}\in\edges\\
  \wedge~ \edge{q}{u}\in\edges
 \end{array} 
 \right\} \\
& & ~~~~~~~~~~~~~~~~~~~~~~~~~~~~~~~~~~~~~~~~~~~~~~~~~~~~~~~~~~~~~~~~~
    if~\edge{p}{q}\in \edges^{2}\nonumber\\
& & 
    \label{eq:asem:4}
  \asubcup{\vcal_{p[1]}} \{ \fbigx_{\edge{q}{u}}|\edge{q}{u}\in\edges \}~
   ~~~~~~~~~~~~~~~~~~~~~~~~~~~~~~~~if~\edge{p}{q}\in \edges^{3}
\end{eqnarray}

\begin{example} {\em Let $P$ the program in example~\ref{ex:program},
    Then $\fsem$ is a system of 23 simultaneous recurrence
    equations.  The following four equations correspond to the four
    equations in example~\ref{ex:collecting} respectively. Let
    $A=member(X,L)$ and $B=member(X,[X|L])$.

\begin{eqnarray*} 
\lefteqn{{[\fsem(\fbigx)]}_{\edge{(3,1)}{(1,1)}} = } \\
 & \aunify{\{X,L,K\}}{\{X,L\}}(A,\fbigx_{\edge{(1,1)}{(5,1)}},B,\asubid{\vcal_{3}})\\
\lefteqn{{[\fsem(\fbigx)]}_{\edge{(5,1)}{(0,0)}} =
\ftheta_{5}}\\ 
\lefteqn{{[
\fsem(\fbigx)]}_{\edge{(1,2)}{(3,1)}} = }\\ 
& \begin{array}{ll}
 &  \aunify{\{X,L\}}{\{X,L,K\}}(B,\fbigx_{\edge{(3,1)}{(1,1)}},
                 A,\fbigx_{\edge{(1,1)}{(5,1)}}) \\
 \asubcup{\{X,L,K\}} &    \aunify{\{X,L\}}{\{X,L,K\}}(B,\fbigx_{\edge{(3,1)}{(1,2)}},
                 A,\fbigx_{\edge{(1,1)}{(5,1)}}) \\
 \asubcup{\{X,L,K\}}  &   \aunify{\{X,L\}}{\{X,L,K\}}(B,\fbigx_{\edge{(3,1)}{(2,1)}},
                 A,\fbigx_{\edge{(1,1)}{(5,1)}}) \\
 \asubcup{\{X,L,K\}}  &   \aunify{\{X,L\}}{\{X,L,K\}}(B,\fbigx_{\edge{(3,1)}{(2,2)}},
                 A,\fbigx_{\edge{(1,1)}{(5,1)}}) \\
 \asubcup{\{X,L,K\}}  &   \aunify{\{X,L\}}{\{X,L,K\}}(B,\fbigx_{\edge{(3,1)}{(4,1)}},
                 A,\fbigx_{\edge{(1,1)}{(5,1)}}) 
  \end{array}\\
\lefteqn{{[ \fsem(\fbigx)]}_{\edge{(1,3)}{(1,2)}} = 
    \fbigx_{\edge{(1,2)}{(3,1)}}~~\asubcup{\{X,L,K\}}~~ \fbigx_{\edge{(1,2)}{(4,2)}}}
\end{eqnarray*} 
$\Box$}
\end{example}

\begin{theorem} \label{th:safeness} {\em
  \(
     \lfp \ssem \ssqsubseteq 
    \fgamma(\lfp \fsem)
  \) if 
\begin{itemize} 
   \item [C3:] \label{scheme:c3} $\epsilon\in
\asubgamma{\vcal}(\asubid{\vcal})$, and
   \item [C4:]
     \(\sunify(A,\asubgamma{\ucal}(\ftheta),
 B,\asubgamma{\vcal}(\fsigma)) \subseteq \asubgamma{\vcal}\fcomp \aunify{\ucal}{\vcal}(A,\ftheta,B,\fsigma)\) for
     any finite $\ucal,\vcal\subseteq\allvars$, any $\ftheta\in
     \asub{\ucal}$, any $\fsigma \in \asub{\vcal}$, and any
     atoms $A$ and $B$ such that $vars(A)\subseteq \ucal$ and
     $vars(B)\subseteq \vcal$. \label{scheme:c4}
 \end{itemize} }

\begin{proof} See p.\pageref{pr:th:safeness}
\end{proof}
\end{theorem} 

\subsection{Complexity of $\fsem$} 
The cost of computing $\lfp \fsem$ is affected by the
characteristics of $P$, the abstract domain, abstract substitutions
$\ftheta_{k}$ where $k\in\aleph_{G}$, and the algorithm for least
fixed-point computation. Using O'Keefe's algorithm for least
fixed-point computation~\cite{OKeefe:87}, the worst case cost of
computing $\lfp \fsem$ is proportional to the product of
the number of operations in $\fsem$ and the maximum height
$dmax$ of $\asub{\vcal_{i}}$ for $i\in\aleph$. Since
$\asubcupf$ is  much less costly than
$\aunifyf$ in most cases, we measure the worst case number of
occurrences of $\aunifyf$ in $\fsem$.

Let $S$ be a set and define $\#S$ be the number of members in $S$. Let
\[\points^{\jmath} \definedas \{ p\in\points~|~\exists q.  \edge{p}{q}\in
\edges^{\jmath}\}\] for $0\leq\jmath\leq 3$.  $\edges^{\jmath}$ is the
set of edges whose ending points lie in $\points^{\jmath}$. Let $pmax$
be the maximum number of predecessors that a program point has.  By
equations~\ref{eq:asem:2} and~\ref{eq:asem:4}, $\aunifyf$ does not
occur in equations for edges in $\edges^{0}\cup\edges^{3}$. By
equation~\ref{eq:asem:1}, $\aunifyf$ occurs at most $pmax$ times in
the equation for an edge in $\edges^{1}$.  So, $\aunifyf$ occurs at
most $\#\edges^{1}*pmax$ times in the equations for edges in
$\edges^{1}$.  By equation~\ref{eq:asem:3}, $\aunifyf$ occurs at
most $pmax^2$ times in the equation for an edge in $\edges^{2}$. So,
$\aunifyf$ occurs at most $\#\edges^{2}*pmax^2$ times in the equations
for edges in $\edges^{2}$. Since $\#\edges^{2}\leq\#\edges^{1}$ and
$\#\edges^{1}\leq \#\aleph_{C}*pmax$, the worst case number of
occurrences of $\aunifyf$ in $\fsem$ is
$O(\#\aleph_{C}*pmax^3)$.  Therefore, the worst case cost of computing
$\lfp \fsem$ is
\[ O(dmax*\#\aleph_{C}*pmax^3)
\]

\section {The Generic Abstract Semantics  $\dsem$} \label{scheme:sec:simpler}

A further approximation may be made of $\fsem$ so as to
reduce the complexity of program analyses.  $\lfp \fsem$ is
a vector indexed by edges in the program graph for $P$. $\lfp
\fsem$ associates with each program point with several abstract
substitutions, each for one edge ending at the program point. This
results in fine analyses for applications such as program debugging.
However, there are some applications where such fine analyses are not
beneficial with respect to their costs and one abstract substitution
for each program point is a better choice.  $\dsem$ fulfills
such purposes and is derived from $\fsem$ by one more
approximation.

The domain of $\dsem$ is $\ddom$ and each
$\dbigx \in \ddom$ is a vector indexed by program points.
$\ddom$ is constructed from $\asub{\vcal}$ in a similar
manner as is $\fdom$.  Let $\dbigx,\dbigy\in
\ddom$, $\dbigx_{p}\in \asub{\vcal_{p[1]}}$ and
define
\begin{eqnarray}
\dbigx \dsqsubseteq \dbigy & \definedas &
  \forall p\in {\cal N}. (\dbigx_{p}~\asuborder{\vcal_{p[1]}} ~\dbigy_{p})\\
{[\dbigx \dsqcap \dbigy]}_{p} &\definedas &
  \dbigx_{p} ~\asubcap{\vcal_{p[1]}}~ \dbigy_{p}\\
{[\dbigx \dsqcup \dbigy]}_{p} &\definedas &
  \dbigx_{p} ~\asubcup{\vcal_{p[1]}}~ \dbigy_{p}\\
\dtop_{p} &\definedas & \asubtop{\vcal_{p[1]}}\\
\dbot_{p} &\definedas & \asubbot{\vcal_{p[1]}}
\end{eqnarray}

The approximation through collapsing the abstract substitutions
associated with all the edges ending at a common program point is
characterised by the following concretization
function.
\begin{equation}
{[\dgamma(\dbigx)]}_{\edge{p}{q}} \definedas  \dbigx_{p}
\label{gamma:collapse} 
\end{equation}
It follows immediately from equation~\ref{gamma:collapse} that
$\dgamma\in \ddom\mapsto \fdom$ is monotonic.
We now construct a monotonic function $\dsem$ on
$<\ddom,\dsqsubseteq>$ such that $ \fsem
\fcomp \dgamma (\dbigx) \fsqsubseteq \dgamma\fcomp
\dsem(\dbigx)$ for every $\dbigx\in \ddom$.

\begin{eqnarray}
\lefteqn{{[\dsem(\dbigx)]}_{p}\definedas} \nonumber\\
& & 
  \ftheta_{p[1]}
  ~~~~~~~~~~~~~~~~~~~~~~~~~~~~~~~~~~~~~~~~~~~~~~~~~~~~~~~~~if~ p\in\points^{0}\label{sem:collapse:2}\\
& & 
   \asubcup{\vcal_{p[1]}} \{ \aunify{\vcal_{q[1]}}{\vcal_{p[1]}}(B_{q},\dbigx_{q},H_{p[1]},
   \asubid{\vcal_{p[1]}}) ~|~
    \edge{p}{q}\in{\edges^{1}}\}\label{sem:collapse:1}\\
& & ~~~~~~~~~~~~~~~~~~~~~~~~~~~~~~~~~~~~~~~~~~~~~~~~~~~~~~~~~~~~~~~if~ p\in\points^{1}\nonumber\\
& & 
  \asubcup{\vcal_{p[1]}} \{
\aunify{\vcal_{q[1]}}{\vcal_{p[1]}}(H_{q[1]},\dbigx_{q},B_{p^{\_}},\dbigx_{p^{\_}})~|~
 \edge{p}{q}\in{\edges^{2}} \}\label{sem:collapse:3}\\
& &
~~~~~~~~~~~~~~~~~~~~~~~~~~~~~~~~~~~~~~~~~~~~~~~~~~~~~~~~~~~~~~~if~p\in\points^{2}\nonumber\\
& & \dbigx_{p^{\_}}
 ~~~~~~~~~~~~~~~~~~~~~~~~~~~~~~~~~~~~~~~~~~~~~~~~~~~~~~~~~~if~p\in\points^{3} \label{sem:collapse:4}
\end{eqnarray}

\begin{lemma}  \label{lm:safeness:simplified} {\em
  \(
     \lfp \fsem  \fsqsubseteq \dgamma( 
    \lfp \dsem)
  \). }

\begin{proof} See p.\pageref{pr:lm:safeness:simplified} 
\end{proof}
\end{lemma} 

$\lfp \dsem$ associates each program point with an
abstract substitution. The simultaneous recurrence equations for
$\dsem$ are simpler than those for $\fsem$ and hence
the computation of $\lfp \dsem$ is less costly than that
of $\lfp \fsem$. $\dsem$ is a
generalisation of Nilsson's generic abstract semantics for definite
logic programs~\cite{Nilsson:88}. Specifically, if $P$ does not have
negative literals then $\lfp \dsem$ is equal to that
in~\cite{Nilsson:88}. Nilsson later~\cite{Nilsson90} presented a
generic abstract semantics that is based on a collecting semantics
that associates with each program point a set of pairs of goal
structures. A goal structure is very similar to a stack in our term.

Each program point in $P$ corresponds to an equation of
$\dsem$. By equations~\ref{sem:collapse:2}
and~\ref{sem:collapse:4}, $\aunifyf$ does not occur in equations
for points in $\points^{0}\cup\points^{3}$. $\aunifyf$ occurs at most
$pmax$ times in an equation for a point in $\points^{1}\cup\points^{2}$
according to equations~\ref{sem:collapse:1} and~\ref{sem:collapse:3}. 
The worst case number of occurrences of $\aunifyf$ in
$\dsem$ is $O(\#\points*pmax)$ since
$\#\points^{1}+\#\points^{2}\leq \#\points$. 
\comments{Note that
$\#\points^{1}=\#\alph_{C}$ is the number of normal clauses in $P$, 
$\#\points^{2}$ is the number of positive literals in the bodies of
normal clauses and queries in $P$.} Therefore, the worst case cost of
computing $\lfp \dsem$  is 
\[ O(dmax*\#\points*pmax)
\] 

$\dsem$ ( a generalisation of the generic abstract semantic
in~\cite{Nilsson:88} ) may also be used to obtain abstract
substitutions for edges. We prefer $\fsem$ to $\dsem$ for such
analyses since $\fsem$ is easier to specialise than $\dsem$.  In
order to specialise $\dsem$ (or the generic abstract semantics
in~\cite{Nilsson90}) for such an analysis, in addition to the work
required to specialise $\fsem$ for the same analysis, one needs to
do
\begin{itemize}
\item keeping information about program points in abstract
  substitutions, an abstract substitution for $\dsem$ is
  a set of  pairs  of a program point and an abstract substitution for
  $\fsem$;
\item replacing $\aunifyf$ by $\dunifyf$ which, for each member of
  $\dtheta\times \dsigma$, discards point information and calls
  $\aunifyf$.
\end{itemize}
This amounts to requiring the analysis design who specialises $\dsem$
to undo approximation $\dgamma$. A call to $\dunifyf$ may cause
$\aunifyf$ to be called as many times as $pmax^2$. So, the worst case
complexity of $\dsem$ is no less than that of $\fsem$ for the same
analysis.

\vspace{1pc}

We have so far developed the generic abstract semantics $\lfp
\fsem$ and $\lfp \dsem$ for forward abstract interpretation of
normal logic programs.  $\lfp \fsem$ obtains an abstract
substitution for each edge in $\edges$ while ${\em lfp} \dsem$
obtains an abstract substitution for each program point in
$\points$. In order to specialise either of these generic abstract
semantics to perform a particular analysis, it is sufficient to
design $\asubf$, $\asubgammaf$, $\asubidf$,$\aunifyf$ and
$\asubcupf$ such that they satisfy C1-C4.

\section {Example} \label{scheme:section:example}

We now illustrate how $\fsem$ and $\dsem$ can be specialised to
perform a particular analysis through groundness analysis - a
simplified version of mode analysis.  

In groundness analysis, we are interested in knowing which variables
will be definitely instantatiated to ground terms. Therefore, a set of
substitutions is approximated naturally by a set of variables.
\(\asub{\vcal}=\wp(\vcal)\).  The partial order on $\asub{\vcal}$
induced by $\subseteq$ on $\wp(Sub)$ is $\supseteq$.
$<\wp(\vcal),\supseteq, \vcal,\emptyset,\cup,\cap>$ is a complete
lattice.

The approximation of a set of substitutions by a set of variables is
modeled by the following concretisation function
$\asubgamma{\vcal}\in \wp(\vcal) \mapsto \wp(Sub)$.
\[
\asubgamma{\vcal} (\ftheta)  \definedas 
  \{ \theta\in Sub~|~\forall X\in \ftheta. (X\theta~is~ground) \}
\]

$\asubgamma{\vcal}$ is obviously a monotonic function from
$<\wp(\vcal),\supseteq>$ to $<\wp(Sub),\subseteq>$ for any
$\vcal\subseteq\allvars$.  For any $\vcal$,
$\asubid{\vcal}=\emptyset$ and $\asubcup{\vcal}=\cap$.

We now present an abstract unification algorithm for groundness
analysis.\linebreak Given $A\in ATOM(\Sigma,\Pi,\ucal)$,
$\ftheta\in \asub{\ucal}$, $B\in ATOM(\Sigma,\Pi,\vcal)$ and
$\fsigma\in \asub{\vcal}$, the algorithm computes
$\aunify{\ucal}{\vcal}(A,\ftheta, B, \fsigma)\in \asub{\vcal}$ in
five steps.  In step (1), a renaming $\Psi$ is applied to $A$ and
$\ftheta$ to obtain $A\Psi$ and $\ftheta\Psi$ so that
$vars(A\Psi)\cap vars(B)=\emptyset$ and $vars(\ftheta\Psi)\cap
vars(\fsigma)=\emptyset$, and $\ftheta\Psi$ and $\fsigma$ are
combined to obtain $\fzeta=\ftheta\Psi\cup\fsigma$ so that a
substitution satisfying $\fzeta$ satisfies both $\ftheta\Psi$ and
$\fsigma$. Note that $\fzeta\in \asub{\ucal\Psi\cup\vcal}$. In
step (2), $E_{0}=mgu(A\Psi,B)$ is computed. If $E_{0}=\fail$ then
the algorithm returns $\asubbot{\vcal}$ that is $\vcal$.
Otherwise, the algorithm continues. In step (3),
$\feta=downwards(E_{0},\fzeta)$ is computed so that $\feta$ is
satisfied by any $\zeta\circ mgu(E_{0}\zeta)$ for any $\zeta$
satisfying $\fzeta$.  In step (4), the algorithm computes $\fbeta
= upwards(\feta,E_{0})$ from $\feta$ such that any substitution
satisfies $\feta$ if it satisfies $\fbeta$ and unifies $E_{0}$. In
step (5), the algorithm restricts $\fbeta$ to $\vcal$ and returns
the result.

\begin{algorithm} \label{groundness:algorithm} 
  {\em Let $\ucal,\vcal\subseteq \allvars$ be finite, $\ftheta\in
    \asub{\ucal}$, $\fsigma\in \asub{\vcal}$, $vars(A)\subseteq
    \ucal$ and $vars(B) \subseteq \vcal$.
\begin{eqnarray*}
  \lefteqn{ \aunify{\ucal}{\vcal}(A,\ftheta,B,\fsigma)
    \definedas} \nonumber\\ &
  \left\{\begin{array}{l}
        let~~~~\Psi~be~a~renaming~such~that~\ucal\Psi\cap\vcal=\emptyset,\\
        ~~~~~~~E_{0}=mgu(A\Psi,B),\\
        if~~~~E_{0}\neq\fail\\
        then~\vcal\cap
        upwards(E_{0},downwards(E_{0},\ftheta\Psi\cup\fsigma))\\
        else~~\vcal
     \end{array}\right. \\
   \lefteqn{downwards(E,\ftheta) \definedas \ftheta \cup
     \bigcup_{(X=t)\in E\wedge X\in\ftheta} vars(t)}\\
  \lefteqn{upwards(E,\ftheta) \definedas
     \ftheta\cup \{X~|~(X=t)\in E\wedge vars(t)\subseteq
     \ftheta\}} 
\end{eqnarray*}  
}
\end{algorithm} 

The abstract domain and the concretisation function satisfy C1-C2
(p.\pageref{scheme:c1c2}) and $\asubid{\vcal}$ satisfies C3. The
following theorem states that algorithm~\ref{groundness:algorithm}
satisfies C4 (p.\pageref{scheme:c4}).

\begin{theorem} \label{groundness:safeness} {\em 
~\begin{itemize} 
\item [C4'] \(
  \sunify(A,\asubgamma{\ucal}(\ftheta),
  B,\asubgamma{\vcal}(\fsigma)) \subseteq
  \asubgamma{\vcal}
  (\aunify{\ucal}{\vcal}(A,\ftheta,B,\fsigma)) \) for any finite
  $\ucal, \vcal\subseteq \allvars$, any $\ftheta\in
  \asub{\ucal}$, any $\fsigma\in \asub{\vcal}$, and any atoms
  $A$ and $B$ such that $vars(A)\subseteq \ucal$ and $vars(B)
  \subseteq \vcal$.
\end{itemize} }

\begin{proof} (C4')
  $\ftheta\Psi\cup\fsigma\in
  \asub{\ucal\Psi\cup\vcal}$.  Let
  $\zeta\in\asubgamma{\ucal\Psi\cup\vcal}
  (\ftheta\Psi\cup\fsigma)$ and\linebreak $Y\in
  downwards(E_{0},\ftheta\Psi\cup\fsigma)$. Then either
  $Y\in \ftheta\Psi\cup\fsigma$ or there is $X$ and $t$ such
  that\linebreak $X\in\ftheta\Psi\cup\fsigma$, $(X=t)\in
  E_{0}$ and $Y\in vars(t)$. So, $Y(\zeta\circ mgu(E_{0}\zeta))$ is
  ground if\linebreak $mgu(E_{0}\zeta)\neq\fail$. It is true that if
  every variable in a term is ground under a substitution then that
  term is ground under the same substitution. Therefore, if\linebreak
  $Z\in upwards(E_{0},
  downwards(E_{0},\ftheta\Psi\cup\fsigma))$ then $Z$ is
  ground under $\zeta\circ mgu(E_{0}\zeta)$. This and
  lemmas~\ref{appendix:lemma:unification} and~\ref{pr:codish} complete
  the proof of C4'.  \end{proof}
\end{theorem} 

\begin{example}{\em~ Let $A=g(U,f(V,f(W,W)),V)$, 
    $B=g(f(X,Y),Z,X)$,$\ftheta=\{U\}$ and\linebreak $\fsigma=\{Z\}$.
$\ftheta$ is an abstract substitution on domain
    $\ucal=\{U,V,W\}$ and $\fsigma$ is an abstract substitution
    on domain $\vcal = \{X,Y,Z\}$.
    This example shows the computation of
    $\aunify{\ucal}{\vcal}(A,\ftheta,B,\fsigma)$ by
    algorithm~\ref{groundness:algorithm}.

    In step (1), a renaming $\Psi=\{U\values U_{0},V\values
    V_{0},W\values W_{0}\}$ is applied to $A$ and $\ftheta$.
        \[      
        \begin{array}{lll} 
                A\Psi &=& g(U_0,f(V_0,f(W_0,W_0)),V_0)\\
                \ftheta\Psi & = & \{U_0\}
        \end{array}
        \] and
        $\fzeta=\ftheta\Psi\cup\fsigma)=\{U_0,Z\}$ is
        computed.

        In step (2), $E_{0} = mgu(A\Psi,B) = \{U_{0}=f(V_0,Y),
        Z=f(V_0,f(W_0,W_0)), X=V_0\}$ is computed. Note that $E_{0}$
        is written as a set of equations in solved form.

        In step (3), $\feta=
        downwards(E_{0},\fzeta)=\{U_0,Z,V_0,Y,W_0\}$ is computed.

        In step (4),
        $\fbeta=upwards(E_0,\feta)=\{U_0,Z,V_0,Y,W_0,X\}$ is
        computed.

        In step (5), the algorithm computes and returns
        $restrict(\fbeta, \vcal)=\{X,Y,Z\}$

        So, $\aunify{\ucal}{\vcal}(A,\ftheta,B,\fsigma) = \{X,Y,Z\}$.
        $\Box$}
\end{example}

\begin{example}
  {\em This example shows the result of the groundness analysis of the
    program in example~\ref{ex:program}. $\lfp\fsem$
    has 23 components each of which corresponds to one edge in
    $\edges$. The following are four of them. 

         \[{[\lfp\fsem]}_{\edge{(3,1)}{(1,1)}} = \{X,L\}\]  
         \[{[\lfp\fsem]}_{\edge{(5,1)}{(0,0)}} = \{Y,Z\}\]  
         \[{[\lfp\fsem]}_{\edge{(1,2)}{(3,1)}} = \{X,L,K\}\] 
	 \[{[\lfp\fsem]}_{\edge{(1,3)}{(1,2)}} = \{X,L,K\}\]
    $\Box$}
\end{example}

\section {Related work and Discussion} \label{scheme:section:related}
\label{scheme:section:discusion}

There has been much research into abstract interpretation of logic
programs.  For a comprehensive survey, see \cite{Cousot:JLP92}. A
number of generic abstract semantics have been brought about for
abstract interpretation of logic
programs~\cite{Bruynooghe91,Kanamori:93,Marriott:JLP92,Mellish:87}.
Abstract interpretation has been used in both forward and backward
analyses of logic programs.  A forward analysis~\cite{Bruynooghe91}
approximates the set of substitutions that might occur at each program
points given a program and a set of goal descriptions. A backward
analysis~\cite{BarbutiGL93,CodishDY94,Marriott:Sondergaard:88,Marriott:JLP92}
approximates the set of the atoms that are logical consequences of a
program~\cite{Emden:Kowalski:76}.  However, the problem of forward
abstract interpretation of {\em normal} logic programs has not been
formally addressed in the literature although negation as failure is
dealt with through the built-in predicate ${!}$ in the way it is
implemented in Prolog. We have proposed a simple solution to the
problem.  We now review previous work and discuss about the solution.

\subsection {Approaches to forward abstract interpretation of logic programs}
There are three approaches to forward abstract interpretation of logic
programs.  A bottom-up forward abstract interpreter mimics a bottom-up
evaluation strategy. A top-down forward abstract interpreter mimics a
top-down evaluation strategy. Top-down forward abstract interpreters
can be further divided into two sub-classes according to whether or
not the underlying top-down evaluation strategy uses memoisation. A
fixed-point forward abstract interpreter computes the least
fixed-point of a system of simultaneous recurrence equations.

\subsubsection{Bottom-up forward abstract interpretation}

The abstract interpreter based on Alexander Templates (AT)
~\cite{Kanamori:93} simulates the bottom-up evaluation based on
AT~\cite{Seki:89}.  Given a program and a goal, AT first transforms
the program and the goal and then evaluates the transformed program
and the transformed goal in a bottom-up manner.  Given a program and a
goal description, the AT-based abstract interpreter  first transforms
the program and the query description in the same way as AT does and
then mimics the evaluation phase of AT by replacing standard
unification with an abstract one.

\subsubsection{Top-down forward abstract interpretation without memoisation}
   
A top-down forward abstract interpreter without
memoisation~\cite{Bruynooghe91,MuthukumarH89a,Waern88,WarrenHD88}
approximately executes a goal description by mimicing the underlying
the top-down evaluation strategy. As an example, we take the top-down
forward abstract interpreter in~\cite{Bruynooghe91}.
    
Given a goal description that is a pair of an atom and an abstract
substitution, the top-down forward abstract interpreter 
in~\cite{Bruynooghe91} constructs an abstract AND-OR graph to
approximate the set of all the intermediate proof trees that may be
constructed by SLDNF under the left-to-right computation rule for all
the goals satisfying the goal description.  In other words, any
intermediate proof tree for any goal satisfying the goal description
can be obtained by unraveling the abstract AND-OR graph. An AND-node
is a clause head and its child OR-nodes are the atoms in the body of
the clause. Every OR-node is adorned with one abstract substitution to
the left, called abstract call substitution, and with another to the
right, called abstract success substitution.  The abstract success
substitution of an OR-node is the abstract call substitution of its
right sibling.

  The initial abstract AND-OR graph has one OR-node that is the atom in
  the goal description and is adorned to the left with the abstract
  substitution in the goal description. Suppose that the abstract
  AND-OR graph has been partly constructed.  Consider an OR-node $A$
  with abstract call substitution $\beta$.  The abstract
  interpreter computes the abstract success substitution of OR-node
  $A$ as follows.  For each clause $C_{i}\equiv H_{i}\leftarrow
  B_{(i,1)},\cdots,B_{(i,m[i])}$ such that $H_{i}$ may match with
  $A\theta$ for some $\theta$ satisfying $\beta$, the abstract
  interpreter  adds to OR-node $A$ a child AND-node $H_{i}$ that has
  $m[i]$ child OR-nodes $B_{(i,1)},\cdots,B_{(i,m[i])}$ and computes
  the abstract call substitution $\beta_{in}^i$ of OR-node $B_{(i,1)}$
  - the first child OR-node of AND-node $H_{i}$.  $\beta_{in}^i$
  approximates the set of the most general unifiers of $H_{i}$ and
  $A\theta$ for all $\theta$ satisfying $\beta$. The abstract
  interpreter  extends OR-node $B_{(i,1)}$ by recursively applying
  the same process and extends OR-node $B_{(i,j+1)}$ in the same way
  after it has computed the abstract success substitution of OR-node
  $B_{(i,j)}$.  Eventually, it will have computed the abstract success
  substitution $\beta_{out}^i$ of OR-node $B_{(i,m[i])}$.  After
  computing $\beta_{out}^i$ for each clause\linebreak $C_{i}\equiv
  H_{i}\leftarrow B_{(i,1)},\cdots,B_{(i,m[i])}$ such that $H_{i}$ may
  match with $A\theta$ for some $\theta$ satisfying $\beta$, the
  abstract interpreter  computes the abstract success substitution
  $\beta'$ of OR-node $A$ and $\beta'$ approximates from above the set
  of the most general unifiers of $A\theta$ and $H_{i}\eta$ for all
  the $\theta$ satisfying $\beta$ and all the $\eta$ satisfying
  $\beta_{out}^i$.

  Since there are recursive calls, \cite{Bruynooghe91} introduces a
  fixed-point component to the abstract interpretation process.
  Suppose that an OR-node $A$ with abstract call substitution $\beta$
  were to be extended. If $A$ has an ancestor OR-node $A'$ with
  abstract call substitution $\beta'$ such that $A$ is a variant of
  $A$ and $\beta$ is a variant of $\beta'$, the abstract
  interpreter  adorns OR-node $A$ to the right with the infimum
  abstract substitution and proceeds until the abstract success
  substitution of OR-node $A'$ is computed. The abstract interpreter 
  then repeatedly recomputes the part of the AND-OR graph starting
  from the abstract success substitution of OR-node $A$ to the
  abstract success substitution of OR-node $A'$ by using the abstract
  success substitution for OR-node $A'$ as the abstract success
  substitution for OR-node $A$. This fixed-point process finishes when
  there is no more increase in the abstract success substitution of
  OR-node $A'$. The same fixed-point component is also used to limit
  the sizes of abstract AND-OR graphs.

  \cite{MuthukumarH89a,Waern88,WarrenHD88} differ from
  \cite{Bruynooghe91} only in dealing with recursive calls.
  ~\cite{MuthukumarH89a} and~\cite{WarrenHD88} make use of a memo
  table and \cite{Waern88} uses stream predicates.

\subsubsection{Top-down forward abstract interpretation with memoisation}

An abstract interpreter  based on an evaluation strategy with
memoisation mimics the underlying evaluation strategy with memoisation
by replacing concrete substitutions with abstract substitutions and
the concrete unification with an abstract unification. For an
introduction to evaluation strategies with memoisation,
see~\cite{Warren92}.

The abstract interpreter based on OLDT
resolution~\cite{KanamoriK87,KanamoriK90,KanamoriJLP93} mimics the
OLDT resolution~\cite{TamakiS86}.  The left-to-right computation rule
is used in OLDT resolution.  Given a goal that is a pair of a sequence
of atoms and a substitution, OLDT resolution~\cite{TamakiS86}
constructs an OLDT structure for the goal.  An OLDT
structure consists of a search tree, a solution table and an
association. An entry of the solution table has a key and a solution
list. The key is an atom and the solution list is a list of atoms that
are instances of the key. Each node of the search tree is a pair of a
goal and a substitution and each edge of the search tree is labeled
with a substitution.  The association is a group of pointers between
the nodes of the search tree and the entries of the solution table.

Initially, the search tree has one node that is the pair of the
sequence of atoms and the substitution, and both the solution table
and the association are empty. OLDT resolution extends the OLDT
structure as follows until it cannot be further extended. Suppose that
the OLDT structure has been partly constructed. Consider a node
$<(A,R),\sigma>$ in the search tree where $A$ is an atom, $R$ a
sequence of atoms and $\sigma$ a substitution. If there is an entry in
the solution table with a key that is a variant of $A\sigma$ then
$<(A,R),\sigma>$ is called a lookup node. Otherwise, it is called a
solution node.  OLDT resolution extends the OLDT structure by
extending its lookup nodes and its leaf solution nodes. If
$<(A,R),\sigma>$ is a leaf solution node then OLDT resolution first
adds into the solution table an entry whose key is $A\sigma$ and whose
solution list is an empty list that will be filled in later.  OLDT
resolution then, for each clause\linebreak $C_{i}\equiv H_{i}\leftarrow
B_{(i,1)},\cdots,B_{(i,m[i])}$ such that $H_{i}$ and $A\sigma$ unify
with $\theta$ being the most general unifier, adds
$<(B_{(i,1)},\cdots,B_{(i,m[i])},R),\sigma\circ\theta>$ as a child
node to node $<(A,R),\sigma>$ and labels the edge from node
$<(A,R),\sigma>$ to node
$<(B_{(i,1)},\cdots,B_{(i,m[i])},R),\sigma\circ\theta>$ with $\theta$.
These child nodes will then be extended by the same process.  If
$<(A,R),\sigma>$ is a lookup node then, OLDT resolution first adds
$<R, \sigma\circ\theta>$ as a child node to $<(A,R),\sigma>$ and
labels the edge from $<(A,R),\sigma>$ to $<R, \sigma\circ\theta>$ with
$\theta$ for each solution $A\sigma\theta$ in the solution list for
key $A\sigma$, and then adds to the association a pointer from lookup
node $<(A,R),\sigma>$ to the tail of the solution list for key
$A\sigma$.  This pointer will be used to add more child nodes to
lookup node $<(A,R),\sigma>$ because at the moment lookup node
$<(A,R),\sigma>$ is first extended, some solutions for $A\sigma$ might
be unavailable from the solution list and will show up later. When a
unit clause is resolved with a leaf solution node $<(A,R),\sigma>$,
the unit clause completes a sub-refutation for $A\sigma$ and may
also completes sub-refutations for the leftmost atoms of other nodes
along the path from $<(A,R),\sigma>$ up to the root of the search
tree. Whenever a unit clause is resolved with a leaf solution node,
OLDT resolution updates the solution lists for those keys that
corresponds to completed sub-resolutions. After the solution list for
a key is updated, OLDT resolution expands those lookup nodes that have
pointers pointing to the solution list accordingly.

The abstract interpreter mimics OLDT resolution closely by
constructing an abstract OLDT structure for a goal description that is
a pair of a sequence of atoms and an abstract substitution.  The nodes
of the abstract OLDT structure are pairs of a sequence of atoms and an
abstract substitution instead of a concrete substitution and the edges
of the abstract OLDT structure are now labeled with abstract
substitutions instead of concrete substitutions. The key of a solution
table entry is now a pair of an atom and an abstract substitution and
so is each solution in the solution list for the key. The abstract
interpreter mimics OLDT by replacing the concrete unification function
with an abstract unification function and the concrete composition
function for concrete substitutions with an abstract composition
function for abstract substitutions.

\subsubsection{Fixed-point forward abstract interpretation} 

Given a program and a set of goal descriptions that are abstract
atoms, \cite{Mellish:87} derives a system of concrete simultaneous
recurrence equations whose least solution approximates the set of all
the input atoms and the set of all the output atoms that occur in an
intermediate proof tree derivable from the program and any goal
satisfying one of the goal descriptions. The system of concrete
simultaneous recurrence equations is approximated from above by a
system of abstract simultaneous recurrence equations with each
concrete operation being replaced by an abstract operation.  Abstract
interpretation is done by computing the least fixed-point of the
system of abstract simultaneous equations.

Given a program and a set of goal descriptions each of which is a pair
of a goal and an abstract substitution, \cite{Nilsson:88} derives a
system of concrete simultaneous recurrence equations whose least
solution gives each program point a superset of the set of all the
possible substitutions at the program point during the satisfaction of
any goal satisfying one of these goal descriptions.  A system of
abstract simultaneous recurrence equations is derived to approximate
from above the system of concrete simultaneous recurrence equations in
the same manner as in~\cite{Mellish:87}.  Abstract interpretation is
accomplished by computing in an abstract domain the least fixed-point
of the system of abstract simultaneous recurrence equations.

\comments{ The collecting semantics that we use is finer
  than~\cite{Nilsson:88} that is finer than~\cite{Mellish:87}. We
  collect sets of substitutions for pairs of program points.
  Specifically, we collect several sets of substitutions for each
  program point, each for one edge ending at the program point.}
\cite{Nilsson:88} collects a set of substitutions for every program
point. ~\cite{Mellish:87} collects the set of input atoms and the set
of output atoms. Collecting the set of input atoms corresponds to
collecting a set of substitutions for the entry point of each clause,
applying each substitution in the set to the head of the clause to
obtain a set of atoms for the clause and then lumping together the
sets of atoms for all the clauses as well as the given set of input
atoms. Similarly, collecting the set of output atoms corresponds to
collecting a set of substitutions for the exit point of each clause,
applying each substitution in the set to the head of the clause to
get a set of output atoms for the clause and lumping together the sets
of output atoms for all the clauses.

\cite{Mellish:87} uses the idea of a trace to summarise the
execution of a query. When making abstraction, the sets of
call substitutions of different calls to the same predicate
are lumped together in a single set input. Similarly, the set
of success substitutions are lumped together in the set
output.

In~\cite{Jones:Sondergaard:87}, contexts are recorded only at the
entry of each program clause.  \cite{Jones:Sondergaard:87} is also a
generic procedure, their core semantics is augmented with application
dependent auxiliary functions that are similar to abstract operations
in \cite{Bruynooghe91}. These auxiliary functions operate on abstract
domains consisting of appropriate approximations of the collecting
semantics. They distinguish between different call instances. However,
there is only one instance of every clause, so substitutions
originating from different call instances are lumped together.

\subsection {The Negation as Failure} 
The treatment of negation as failure in $\fsem$ and $\dsem$ is
simple. The transition system $\sem$ approximates VSLDNF (an
equivalent of SLDNF) by assuming that a negative literal always
succeeds while it may fail. This approximation introduces noises into
$\fsem$ and $\dsem$.  However, it is difficult within the provisions
of abstract interpretation to improve on this simple solution.

Let $(\neg A)\sigma$ be selected by SLDNF where $\neg A$ is a
negative literal in the body of a clause in the program and $\sigma$
be a substitution. During abstract interpretation, $\sigma$ is not
known and possible values for $\sigma$ are described by an abstract
substitution $\fsigma\in \asub{\vcal}$, often called the
abstract call substitution for $(\neg A)$.  Since $\fsigma$
usually describes an infinite set of substitutions, it may well be the
case that $A\sigma$ succeeds for some $\sigma\in
\asubgamma{\vcal}(\fsigma)$ and fails for other
$\sigma\in \asubgamma{\vcal}(\fsigma)$. We take
$\fsigma$ as the abstract success substitution for $\neg A$ by
simply assuming that $A\sigma$ fails for all $\sigma\in
\asubgamma{\vcal}(\fsigma)$ ($\neg A\sigma$
succeeds). An improvement needs making the abstract success
substitution for $\neg A$ stronger, that is, replacing $\fsigma$
with another abstract substitution $\feta\in \asub{\vcal}$
such that $\feta ~\asuborder{\vcal}~
\fsigma$. Let us assume that
$\asubgamma{\vcal}(\fsigma)\backslash
\asubgamma{\vcal}(\eta) \neq\emptyset$ for otherwise
$\feta$ is no stronger than $\fsigma$. By safeness
requirement for negation as failure and safeness requirement for
abstract interpretation, it is necessary to be able to infer
\begin{equation} \label{eq:trouble}
  \forall \theta\in
  \asubgamma{\vcal}(\fsigma)\backslash
  \asubgamma{\vcal}(\eta). \left(
\begin{array}{l}
  \epsilon~is~a~computed~answer~for~P\cup\{\leftarrow A\theta\}
\end{array}\right)
\end{equation}
To infer~\ref{eq:trouble}, we need to {\em under-estimate} success
and {\em over-estimate} failure in order to make the analysis safe.
However, abstract interpreters {\em over-estimate} success and {\em
  under-estimate} failure.  Note that the word approximation in
abstract interpretation means approximation {\em from above}. An
abstract interpreter {\em over-estimates} success by means of an
abstract unification function which approximates the normal
unification function {\em from above}, that is, {\em over-estimates}
the success of the normal unification function. To
infer~\ref{eq:trouble}, we must use a unification function which
approximates the normal unification function {\em from below}. Such a
unification function should succeed only if the normal unification
function succeeds. Of course, we could use a unification function
which always fails. But, this does not achieve any improvement since
such a unification function will make $A\sigma$ always fail.  Since an
abstract domain for an analysis is much simpler than the concrete
domain, some information about a set of substitutions is lost when the
set of substitutions is approximated by an abstract substitution.  It
is difficult to design a unification function which approximates the
normal unification function {\em from below} based on abstract
substitutions because abstract substitutions are inaccurate
descriptions of sets of substitutions while such a unification
function, we believe, needs accurate descriptions of sets of
substitutions.

\comments{
\subsection {Modification of Existing Generic Abstract Semantics}

We have derived $\fsem$ and $\dsem$ though a series of
approximations of operational semantics SLDNF.  The generic abstract
semantics for forward abstract interpretation of logic programs in the
literature~\cite{Bruynooghe91,Kanamori:93,Mellish:87} are derived by
means of a series of approximations of operational semantics SLD. We
believe that these generic abstract semantics can be reworked by using
SLDNF instead of SLD as operational semantics and introducing an extra
approximation which approximates SLDNF by a transition system as we
have done. The derivation and the safeness proof of the resultant
generic abstract semantics could look like a standard execise and the
resultant generic abstract semantics could be intimate to the original
generic abstract semantics.  For instance,
$\dsem$ is intimate to that in~\cite{Nilsson:88}.  However,
it is essential to do such exercises rather than making changes to the
original generic abstract semantics derived from SLD and
opportunistically assuming them to be safe. A generic abstract
semantics without a sound safeness proof cannot be used with
confidence.

\subsection {Direct Use of Generic Abstract Semantics}

Another plausible solution to the problem of forward abstract
interpretation of normal logic programs is first tranform a normal
logic program into a definite logic program and then directly apply an
existing generic abstract semantics to the definite logic program.
Consider the following transformation.  Let $P$ be the normal logic
program and $P'$ the definite logic program. $P'$ is obtained from $P$
by changing each $L_{p} \equiv \neg B_{p} \equiv \neg a(\bar{X})$ into
$not\_a(\bar{X})$ where $\bar{X}$ is a vector of variables occuring in
$L_{p}$. In addition, $P'$ contains, for each $L_{p} \equiv \neg B_{p}
\equiv \neg a(\bar{X})$,
\[\begin{array}{l}  not\_a(\bar{X})\leftarrow a(\bar{X}),fail\\
                    not\_a(\bar{X})
  \end{array} \]
  There may be many ther transformations. Let $P'$ be
  obtained from $P$ though the above (or any other) transforamtion.
  Before analysing $P$ by applying an existing generic abstract
  semantics  to analyse $P'$, we need a proof for the following. 
\begin{quotation}  For any $\leftarrow\Vector{L}{m}$, if an
  SLDNF-derivation of $P\cup\{\leftarrow\Vector{L}{m}\}$ reaches a
  program point $p$ with a current substitution $\sigma$ then there is
  an SLD-derivation of $P'\cup\{\leftarrow\Vector{L}{m}\}$ that
  reaches $p$ with a current substitution $\sigma'$ such that
  $X\sigma'$ and $X\sigma$ are equivalent (module renaming) for any
  variable $X$ in the clause (or the query) of $p$.
\end{quotation} 
We do not know if there is a proof for the above hypothesis. Before
a proof is provided,  the safeness of the analysis cannot be
guaranteed. 
}

\section {Summary}  \label{scheme:section:conclusion} 
We have presented and justified a simple solution to the problem of
forward abstract interpretation of normal logic programs
and derived  generic abstract semantics $\lfp
\fsem$ and $\lfp \dsem$ of normal logic
programs. The solution is simple and it amounts to replacing the
negation as failure rule with an unconditional derivation rule. 

$\lfp \fsem$ and $\lfp \dsem$ can be
specialised for various analyses. An analysis can be thought of as a
series of approximations of the operational semantics as shown below.
An arrow from $A$ to $B$ reads as ``A is approximated by B''.

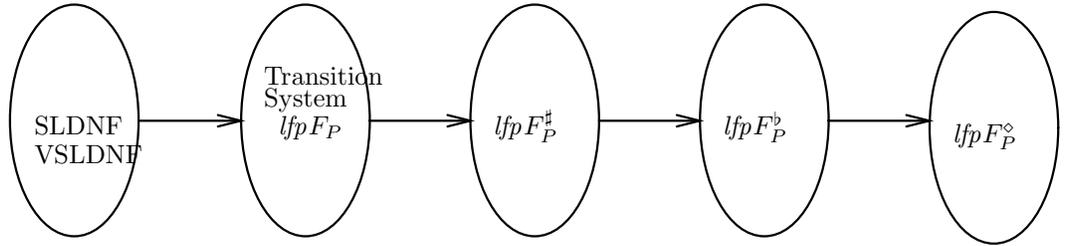
\begin{center} 
\setlength{\unitlength}{0.0075in}
\begin{picture}(731,182)(0,-10)
\thicklines
\put(686,81){\ellipse{90}{162}}
\put(651,71){\makebox(0,0)[lb]{\raisebox{0pt}[0pt][0pt]{\shortstack[l]{{ $\lfp \dsem$}}}}}
\put(45,86){\ellipse{90}{162}}
\put(206,86){\ellipse{90}{162}}
\put(366,86){\ellipse{90}{162}}
\put(526,86){\ellipse{90}{162}}
\path(91,86)(161,86)
\path(145.000,82.000)(161.000,86.000)(145.000,90.000)
\path(251,86)(321,86)
\path(305.000,82.000)(321.000,86.000)(305.000,90.000)
\path(411,86)(481,86)
\path(465.000,82.000)(481.000,86.000)(465.000,90.000)
\path(571,86)(641,86)
\path(625.000,82.000)(641.000,86.000)(625.000,90.000)
\put(171,111){\makebox(0,0)[lb]{\raisebox{0pt}[0pt][0pt]{\shortstack[l]{{ Transition}}}}}
\put(171,96){\makebox(0,0)[lb]{\raisebox{0pt}[0pt][0pt]{\shortstack[l]{{ System}}}}}
\put(181,76){\makebox(0,0)[lb]{\raisebox{0pt}[0pt][0pt]{\shortstack[l]{{ $\lfp \sem$}}}}}
\put(331,76){\makebox(0,0)[lb]{\raisebox{0pt}[0pt][0pt]{\shortstack[l]{{ $\lfp \ssem$}}}}}
\put(491,76){\makebox(0,0)[lb]{\raisebox{0pt}[0pt][0pt]{\shortstack[l]{{ $\lfp \fsem$}}}}}
\put(11,76){\makebox(0,0)[lb]{\raisebox{0pt}[0pt][0pt]{\shortstack[l]{{ SLDNF}}}}}
\put(11,56){\makebox(0,0)[lb]{\raisebox{0pt}[0pt][0pt]{\shortstack[l]{{ VSLDNF}}}}}
\end{picture}

\end{center} 

To specialise either $\lfp \fsem$ or $\lfp \dsem$ for an analysis, one
has to find out the corresponding abstract domain and corresponding
concretisation function, to provide a function for computing abstract
identity substitutions, a function for computing the least upper
bounds, and a function for computing abstract unifications. The
abstract domain and the concretisation function must satisfy C1-C2,
the function for computing abstract identity substitutions must
satisfy C3, and the function for computing abstract unifications must
satisfy C4.

We deal with negation as failure by approximating SLDNF with a
transition system. The way negation as failure is dealt with may be
generalised to deal with built-in predicate $!$. Although, we have not
dealt with other built-in predicates, we believe that the generic
abstract semantics can be augmented to deal with these built-in
predicates in the way they are dealt with in~\cite{Bruynooghe91}.

\section{Appendix}

\begin{lemma} \label{appendix:lemma:unification}
{\em  Let $\rho$ be a renaming such that $vars(a\rho)\cap
  vars(b)=\emptyset$ and $vars(\phi\rho)\cap vars(\psi)=\emptyset$. If
  $(a\rho)(\phi\rho)$ and $b\psi$ unify then $a\rho$ and $b$ unify.}

\begin{proof} Let $a\rho\equiv a'$ and $\phi\rho\equiv \phi'$. If
    $a'\phi'$ and $b\psi$ unify then there is a substitution $\theta$
    such that $a'\phi'\theta\equiv b\phi\theta$.  We have
    $vars(a')\cap dom(\psi)=\emptyset$ and $rang(\psi)\cap
    dom(\phi')=\emptyset$ and $vars(b)\cap dom(\phi')=\emptyset$.
    Hence, $a'\psi\phi'\theta\equiv b\psi\phi'\theta$.  Therefore,
    $a\rho$ and $b$ unify.  \end{proof}
\end{lemma}

\begin{lemma} \label{appendix:lemma:renaming}
{\em  Let $A$ and $B$ be two atoms, and $\rho_{1}$ and $\rho_{2}$ be two
  renamings such that
\begin{eqnarray} 
 \label{alu:loc1} dom(\rho_{1})=dom(\rho_{2})\supseteq vars(B)\\ 
 \label{alu:loc2} rang(\rho_{1})\cap vars(A)=\emptyset\\ 
 \label{alu:loc3} rang(\rho_{2})\cap vars(A)=\emptyset
\end{eqnarray} 
Then
\begin{itemize}
\item [(a)] $A$ and $B\rho_{1}$ unify iff  $A$ and $B\rho_{2}$
unify. 
\item [(b)] $mgu(A,B\rho_{1})\uparrow vars(A)\cong
  mgu(A,B\rho_{2})\uparrow vars(A)$.
\item [(c)] \(\rho_{1}\circ mgu(A,B\rho_{1})\uparrow dom(\rho_{1}) 
        \cong\rho_{2}\circ mgu(A,B\rho_{2})\uparrow dom(\rho_{2})\). 
\end{itemize} }

\begin{proof} Let $vars(A)=\{X_{1},\cdots,X_{k}\}$,
$dom(\rho_{1})=dom(\rho_{2})=\{V_{1},\cdots,V_{l}\}$,\linebreak
$\rho_{1}=\{V_{1}\values Y_{1},\cdots,V_{l}\values Y_{l}\}$ and 
$\rho_{2}=\{V_{1}\values Z_{1},\cdots,V_{l}\values
Z_{l}\}$.  Define \[\rho_{3} \definedas \{ Z_{1}\values Y_{1},\cdots,
Z_{l}\values Y_{l} \}\]
\[\rho_{4} \definedas \{Y_{1} \values Z_{1},
\cdots,  Y_{l} \values Z_{l} \}\]
\[ \ycal \definedas  \{Y_{1},\cdots,Y_{l} \}
\]
\[ \zcal  \definedas  \{Z_{1},\cdots,Z_{l}\}
\] 
\[
   \vcal \definedas  \{V_{1},\cdots,V_{l}\}
\]
We have
\begin{equation} \label{alu:loc4} 
   \rho_{1}=\rho_{2}\circ \rho_{3} \uparrow \vcal
\end{equation} 
and \begin{equation} \label{alu:loc5}
    \rho_{2}=\rho_{1}\circ \rho_{4} \uparrow \vcal
\end{equation}

Suppose that $A$ and $B\rho_{1}$ unify with $\theta_{1}$ being their most
general unifier. Let
\begin{equation} \label{alu:loc6} 
  \theta_{1} = \{ X_{i_{1}}\values x_{i_{1}},\cdots,
  X_{i_{s}}\values x_{i_{s}}, Y_{j_{1}}\values
      y_{j_{1}},\cdots,Y_{j_{t}}\values y_{j_{t}} \}
\end{equation} 
with $1\leq i_{1}\leq\cdots\leq i_{s}\leq k$ and $1\leq
j_{1}\leq\cdots\leq j_{t}\leq l$.  Define 
\begin{equation} \label{alu:loc7}
   y_{h}\definedas  \left\{ \begin{array}{ll} 
               Y_{h} & If~h\not\in \{j_{1},j_{2},\cdots,j_{t}\}\\
               y_{h} & If~h\in \{j_{1},j_{2},\cdots,j_{t}\}
           \end{array}\right.
\end{equation} By equations~\ref{alu:loc6}-\ref{alu:loc7},  
we have 
\begin{equation} \label{alu:loc8}
 \rho_{1}\circ \theta_{1} \uparrow \vcal = \{V_{1}\values
y_{1},\cdots,V_{l}\values y_{l} \} 
\end{equation} 

$ A\rho_{3}\theta_{1} =
A\theta_{1}=B\rho_{1}\theta_{1}=B(\rho_{2}\circ\rho_{3}\uparrow
\vcal)\theta_{1} = B\rho_{2}\rho_{3}\theta_{1}$ by
equations~\ref{alu:loc1},~\ref{alu:loc3},~\ref{alu:loc4},~\ref{alu:loc6}
and~\ref{alu:loc7}.  So, $A$ and $B\rho_{2}$ unify with
$\rho_{3}\theta_{1}$ being one of their unifiers if $A$ and
$B\rho_{1}$ unify with $\theta_{1}$ being their most general unifier.

Suppose $A$ and $B\rho_{2}$ unify with $\theta_{2}$ being their most
general unifier. Let
\begin{equation} \label{alu:loc9} 
  \theta_{2} =
  \{X_{u_{1}} \values
  \overline{x}_{u_{1}},\cdots,X_{u_{p}}\values \overline{x}_{u_{p}},
  Z_{v_{1}}\values {z}_{v_{1}},\cdots, {Z}_{v_{q}} \values 
  {z}_{v_{q}} \}
\end{equation} 
with $1\leq u_{1}\leq\cdots\leq u_{p}\leq k$ and 
$1\leq v_{1}\leq \cdots\leq v_{q}\leq l$. Define 
\begin{equation} \label{alu:loc10} 
   z_{h}\definedas  \left\{ \begin{array}{ll}
               Z_{h} & If~h\not\in \{v_{1},v_{2},\cdots,v_{q}\}\\
               z_{h} & If~h\in \{v_{1},v_{2},\cdots,v_{q}\}
           \end{array}\right.
\end{equation} By equations~\ref{alu:loc9}-\ref{alu:loc10},
we have
\begin{equation} \label{alu:loc11}
\rho_{2}\circ \theta_{2}\uparrow \vcal = \{V_{1}\values
z_{1},\cdots,V_{l}\values z_{l} \}
\end{equation}

$ A\rho_{4}\theta_{2} =
A\theta_{2}=B\rho_{2}\theta_{2}=B(\rho_{1}\circ\rho_{4}\uparrow
\vcal)\theta_{2} = B\rho_{1}\rho_{4}\theta_{2}$ by
equations~\ref{alu:loc1}-\ref{alu:loc2}, \ref{alu:loc5},
and~\ref{alu:loc9}-\ref{alu:loc10}. So, $A$ and $B\rho_{1}$ unify with
$\rho_{4}\theta_{2}$ being one of their unifiers if $A$ and
$B\rho_{2}$ unify with $\theta_{2}$ being their most general unifier.
Therefore, (a) holds.

The following equation results from equation~\ref{alu:loc6} and the
definition of $\rho_{3}$. 

\begin{equation}\label{alu:loc12} 
\rho_{3}\theta_{1} = \left(\begin{array}{c}
  \{ X_{i_{1}}\values x_{i_{1}},\cdots,
  X_{i_{s}}\values x_{i_{s}}\}\\
 \cup \\
 \{Y_{j_{o}}\values y_{j_{o}}~|
    ~1\leq o\leq t\wedge Y_{j_{o}}\not\in\zcal\} \\
   \cup \\ \{Z_{1}\values y_{1},\cdots,Z_{l}\values y_{l}\} \end{array}
 \right)
\end{equation} 

The following equation results from equation~\ref{alu:loc9} and the
definition of $\rho_{4}$. 

\begin{equation}\label{alu:loc13}
 \rho_{4}\theta_{2} = \left(\begin{array}{c}
   \{X_{u_{1}} \values
  \overline{x}_{u_{1}},\cdots,X_{u_{p}}\values \overline{x}_{u_{p}} \}\\
  \cup \\
  \{ Z_{v_{o}}\values {z}_{v_{o}}~|~1\leq o\leq q\wedge
  {Z}_{v_{o}}\not\in\ycal \} \\ 
  \cup \\ \{Y_{1}\values z_{1},\cdots,Y_{l}\values z_{l} \}
\end{array} \right)
\end{equation} 

Since $\rho_{4}\theta_{2}$ is a unifier of $A$ and $B\rho_{1}$, there
is a substitution $\delta_{1}$ such that
$\rho_{4}\theta_{2}=\theta_{1}\delta_{1}$. By equations~\ref{alu:loc6}
and~\ref{alu:loc13}, we have 

\begin{eqnarray} \label{alu:loc14} 
\lefteqn{ \left(\begin{array}{c}
  \{X_{u_{1}} \values
  \overline{x}_{u_{1}},\cdots,X_{u_{p}}\values \overline{x}_{u_{p}} \}\\
  \cup \\
    \{ Z_{v_{o}}\values {z}_{v_{o}}~|~1\leq o\leq q\wedge
  {Z}_{v_{o}}\not\in \ycal 
  \}\\ \cup \\ \{Y_{1}\values z_{1},\cdots,Y_{l}\values z_{l} \}
  \end{array}\right)
= }  \\ 
 & \left( \begin{array}{c}
  \{ X_{i_{1}}\values x_{i_{1}},\cdots,
  X_{i_{s}}\values x_{i_{s}}\}\\ \cup \\
  \{Y_{j_{1}}\values
      y_{j_{1}},\cdots,Y_{j_{t}}\values y_{j_{t}} \}\end{array}\right
      )\delta_{1} \nonumber
\end{eqnarray} 

Since $\rho_{3}\theta_{1}$ is a unifier of $A$ and $B\rho_{2}$, there
is a substitution $\delta_{2}$ such that
$\rho_{3}\theta_{1}=\theta_{2}\delta_{2}$. By 
 equations~\ref{alu:loc9}
and~\ref{alu:loc12}, we have 
\begin{eqnarray} \label{alu:loc15} 
\lefteqn{
\left(\begin{array}{c} 
  \{ X_{i_{1}}\values x_{i_{1}},\cdots, X_{i_{s}}\values x_{i_{s}}\}\\
  \cup \\
 \{Y_{j_{o}}\values y_{j_{o}}~|
    ~1\leq o\leq t\wedge Y_{j_{o}}\not\in\zcal \} \\
  \cup\\
  \{Z_{1}\values y_{1},\cdots,Z_{l}\values y_{l}\}
 \end{array}\right)
=}\\
& \left( \begin{array}{c}
   \{X_{u_{1}}\values\overline{x}_{u_{1}},\cdots,X_{u_{p}}\values
      \overline{x}_{u_{p}}\}\\
    \cup \\ \{Z_{v_{1}}\values {z}_{v_{1}},\cdots,{Z}_{v_{q}}\values
            {z}_{v_{q}}\}
  \end{array}\right)\delta_{2} \nonumber
\end{eqnarray} 

By equation~\ref{alu:loc14}, $\{X_{i_{1}},\cdots,X_{i_{s}}\}\subseteq
\{ X_{u_{1}},\cdots,X_{u_{p}} \}$ and, by equation~\ref{alu:loc15},\linebreak
$\{ X_{u_{1}},\cdots,X_{u_{p}} \} \subseteq
\{X_{i_{1}},\cdots,X_{i_{s}}\}$. So,
$\{X_{i_{1}},\cdots,X_{i_{s}}\}=\{ X_{u_{1}},\cdots,X_{u_{p}} \} $. We
have $s=p$ and $i_{o}=u_{o}$ for $1\leq o\leq s$. We also have, from
equations~\ref{alu:loc14}-\ref{alu:loc15},
\begin{equation} \label{alu:loc16} 
\begin{array}{l}
 x_{i_{o}} = \overline{x}_{i_{o}}\delta_{2}\\
 \overline{x}_{i_{o}} = x_{i_{o}}\delta_{1}
\end{array} 
\end{equation}  
$x_{i_{o}}\cong\overline{x}_{i_{o}}$ for $1\leq o\leq s$ from
equation~\ref{alu:loc16}. Therefore, (b) holds.

By equation~\ref{alu:loc14}, we have 
\begin{eqnarray} \label{alu:loc17}
Y_{h}\values z_{h}\in\delta_{1} & If~h\not\in\{j_{1},\cdots,j_{t}\}\\ 
\label{alu:loc18}
z_{h}=y_{h}\delta_{1} & If~h\in\{j_{1},\cdots,j_{t}\}
\end{eqnarray} 

$y_{h}=Y_{h}$ if $h\not\in\{j_{1},\cdots,j_{t}\}$ by
equation~\ref{alu:loc7}. So, $z_{h}=y_{h}\delta_{1}$ for
$h\not\in\{j_{1},\cdots,j_{t}\}$ by equation~\ref{alu:loc17}. This and
equation~\ref{alu:loc18} imply that, for all $1\leq h\leq l$,
\begin{equation} \label{alu:loc19} 
z_{h}=y_{h}\delta_{1} 
\end{equation} 

By equation~\ref{alu:loc15}, we have
\begin{eqnarray} \label{alu:loc20}
Z_{h}\values y_{h}\in\delta_{2} &
If~h\not\in\{v_{1},\cdots,v_{q}\}\\ 
\label{alu:loc21}
y_{h}=z_{h}\delta_{2} & If~h\in\{v_{1},\cdots,v_{q}\}
\end{eqnarray}

$z_{h}=Z_{h}$ if $h\not\in\{v_{1},\cdots,v_{q}\}$ by
equation~\ref{alu:loc10}. So, $y_{h}=z_{h}\delta_{2}$ for
$h\not\in\{v_{1},\cdots,v_{q}\}$ by equation~\ref{alu:loc20}. This  and
equation~\ref{alu:loc21} imply that for all $1\leq h\leq l$
\begin{equation} \label{alu:loc22} 
y_{h}=z_{h}\delta_{2}
\end{equation} 

By equations~\ref{alu:loc8},~\ref{alu:loc11},~\ref{alu:loc19}
and~\ref{alu:loc22}, $\rho_{1}\circ\theta_{1}\uparrow
\vcal\cong\rho_{2}\circ\theta_{2}\uparrow \vcal$.
Therefore, (c) holds. \end{proof}
\end{lemma}

\begin{corollary} \label{appendix:corollary:unification:2}
{\em Let $A$ and $B$ be two atoms and $\rho$ be a renaming
such that $dom(\rho)\supseteq vars(B)$. If
\( vars(A)\cap vars(B)=\emptyset
\) and 
\( vars(A)\cap vars(B\rho)=\emptyset
\) then $A$ and $B$ unify iff $A$ and $B\rho$ unify, and
\[mgu(A,B)\uparrow vars(B)\cong (\rho\circ mgu(A,B\rho))\uparrow
vars(B)\]}

\begin{proof} The proof results immediately from
  lemma~\ref{appendix:lemma:unification}.(a) and (c) by letting
  $\rho_{2}=\rho$ and $\rho_{1}$ be the renaming on $vars(B)$ such
  that $X\rho_{1}=X$ for each $X\in vars(B)$.
\end{proof}
\end{corollary}

\begin{corollary} \label{appendix:corollary:unification:3} 
{\em  Let $A$ and $B$ be two atoms, $\rho_{A}$ and $\rho_{B}$ be renamings. If
\[dom(\rho_{A})\supseteq vars(A)\]
\[dom(\rho_{B})\supseteq vars(B)\]
\[ vars(A\rho_{A}) \cap vars(B) = \emptyset
\]
\[ vars(B\rho_{B}) \cap vars(A) = \emptyset
\] then $A\rho_{A}$ and $B$ unify iff $A$ and $B\rho_{B}$ unify, and
\[(\rho_{A}\circ mgu(A\rho_{A},B))\uparrow dom(\rho_{A})\cong
mgu(A,B\rho_{B})\uparrow vars(A)\]}

\begin{proof} We prove the {\em if} part. The {\em only if} part is a dual case
  of the {\em if} part.  Let $\rho_{B}'$ be a renaming such that
  $dom(\rho_{B}')=dom(\rho_{B})$, $vars(B\rho_{B}')\cap
  vars(A)=\emptyset$ and $vars(A\rho_{A})\cap
  vars(B\rho_{B}')=\emptyset$.

  Suppose that $A$ and $B\rho_{B}$ unify. By
  lemma~\ref{appendix:lemma:unification}.(a), $A$ and $B\rho_{B}'$
  unify, and $mgu(A,B\rho_{B}') \uparrow vars(A)\cong
  mgu(A,B\rho_{B})\uparrow vars(A)$ by
  lemma~\ref{appendix:lemma:unification}.(b).  By
  corollary~\ref{appendix:corollary:unification:2}, $A\rho_{A}$ and
  $B\rho_{B}'$ unify, and \[\rho_{A}\circ
  mgu(A\rho_{A},B\rho_{B}')\uparrow vars(A) \cong
  mgu(A,B\rho_{B}')\uparrow vars(A)\] So, $\rho_{A}\circ
  mgu(A\rho_{A},B\rho_{B}')\uparrow vars(A) \cong
  mgu(A,B\rho_{B})\uparrow vars(A)$.  By
  corollary~\ref{appendix:corollary:unification:2}, $A\rho_{A}$ and
  $B$ unify, and \[mgu(A\rho_{A},B)\uparrow vars(A\rho_{A})\cong
  mgu(A\rho_{A},B\rho_{B}')\uparrow vars(A\rho_{A})\] hence
  $\rho_{A}\circ mgu(A\rho_{A},B)\uparrow vars(A)\cong \rho_{A}\circ
  mgu(A\rho_{A},B\rho_{B}')\uparrow vars(A)$.  Therefore,
  $\rho_{A}\circ mgu(A\rho_{A},B)\uparrow vars(A) \cong
  mgu(A,B\rho_{B})\uparrow vars(A)$.
  It now suffices to prove \(\rho_{A}\circ mgu(A\rho_{A},B)\uparrow
  dom(\rho_{A}) \cong \rho_{A}\circ mgu(A\rho_{A},B)\uparrow vars(A)\).
  Let \( \rho_{A}^1 = \rho_{A} \uparrow vars(A)\) and 
      \( \rho_{A}^2 = \rho_{A} \uparrow (dom(\rho_{A})-vars(A)) \). 
We have $\rho_{A}=\rho_{A}^1\cup\rho_{A}^2$,
\begin{eqnarray*} 
 \lefteqn{\rho_{A}\circ mgu(A\rho_{A},B)\uparrow dom(\rho_{A}) } \\
   & \begin{array} {l} = (\rho_{A}^1\cup\rho_{A}^2)\circ
             mgu(A(\rho_{A}^1\cup\rho_{A}^2),B) \uparrow dom(\rho_{A}) \\
             =  \rho_{A}^1\circ mgu(A\rho_{A}^1,B) \uparrow vars(A)
             \cup  \rho_{A}^2 
      \end{array}
  \end{eqnarray*}
and 
 \begin{eqnarray*} 
 \lefteqn{\rho_{A}\circ mgu(A\rho_{A},B)\uparrow vars(A)}\\
   & \begin{array} {l} = (\rho_{A}^1\cup\rho_{A}^2)\circ
          mgu(A(\rho_{A}^1\cup\rho_{A}^2),B) \uparrow vars(A) \\
          =  \rho_{A}^1\circ mgu(A\rho_{A}^1,B) \uparrow vars(A)
     \end{array}
  \end{eqnarray*}

We also have 
$rang(\rho_{A}^1\circ mgu(A\rho_{A}^1,B) \uparrow vars(A))\cap
dom(\rho_{A}^2)=\emptyset$ and\linebreak $dom(\rho_{A}^2)\cap vars(A)=\emptyset$. So, 
\begin{eqnarray*}
\lefteqn{(\rho_{A}\circ mgu(A\rho_{A},B) \uparrow vars(A)) \circ
\rho_{A}^2}\\
& \begin{array} {l}
   = (\rho_{A}^1\circ mgu(A\rho_{A}^1,B) \uparrow vars(A))\circ \rho_{A}^2\\
   = \rho_{A}^1\circ mgu(A\rho_{A}^1,B) \uparrow vars(A) \cup \rho_{A}^2\\
   = \rho_{A}\circ mgu(A\rho_{A},B)\uparrow dom(\rho_{A})
  \end{array}
\end{eqnarray*}
Therefore, $\rho_{A}\circ mgu(A\rho_{A},B)\uparrow
dom(\rho_{A})\cong\rho_{A}\circ mgu(A\rho_{A},B)\uparrow vars(A)$
since $\rho_{A}^2$ is a renaming.
\end{proof}
\end{corollary}

\begin{lemma} \label{appendix:lemma:composition:restriction} Let
  {\em $\theta_{1}$ and $\theta_{2}$ be two substitutions and \vcal a
    set of variables.
  \[ 
        \theta_{1}\circ\theta_{2}\uparrow \vcal =
        (\theta_{1}\uparrow \vcal)\circ \theta_{2}\uparrow \vcal
  \]}

\begin{proof} Let $(X \values t)\in\theta_{1}\circ\theta_{2}\uparrow \vcal$.
    Then $X\in\vcal$. Either $X\in dom(\theta_{1})$ or $X\not\in
    dom(\theta_{1}) \wedge X\in dom(\theta_{2})$.  If $X\in
    dom(\theta_{1})$ then there is $t_{1}$ such that $((X\values t_{1})\in\theta_{1}\wedge
    t=t_{1}\theta_{2})$. Since $X\in\vcal$,
    $(X\values t_{1})\in\theta_{1}\uparrow\vcal$ and hence
    $X\values (t_{1}\theta_{2}) = (X\values t)\in(\theta_{1}\uparrow \vcal)\circ
    \theta_{2}\uparrow \vcal$. Otherwise, $X\in dom(\theta_{2})$,
    $(X\values t)\in\theta_{2}$ and $(X\values t)\in(\theta_{1}\uparrow \vcal)\circ \theta_{2}\uparrow \vcal$.

    Let $(X\values t)\in (\theta_{1}\uparrow \vcal)\circ
    \theta_{2}\uparrow \vcal$. Then $X\in\vcal$. Either $X\in
    dom(\theta_{1}\uparrow \vcal)$ or $X\not\in \theta_{1}\uparrow
    \vcal\wedge X\in dom(\theta_{2})$. If $X\in dom(\theta_{1}\uparrow
    \vcal)$ then there is $t_{2}$ such that $((X\values
    t_{2})\in\theta_{1}\uparrow\vcal\wedge t=t_{2}\theta_{2})$.
    $(X\values t_{2})\in\theta_{1}$ and $(X\values
    t)\in\theta_{1}\circ\theta_{2}$. So, $(X\values
    t)\in\theta_{1}\circ\theta_{2}\uparrow \vcal$. Otherwise,
    $(X\values t)\in\theta_{2}$ and $X\not\in
    dom(\theta_{1})\cap\vcal$.  So, $(X\values
    t)\in\theta_{1}\circ\theta_{2}\uparrow \vcal$. 
\end{proof}
\end{lemma}

\begin{lemma} \label{pr:codish} 
  {\em Let $E_{1}$ and $E_{2}$ be two sets of equations, and $\theta_{1}$
    and $\theta_{2}$ be two substitutions.  If $\theta_{1} =
    mgu(E_{1})$ and $\theta_{2}= mgu(E_{2}\theta_{1})$ then
    $\theta_{1}\circ\theta_{2}= mgu(E_{1}\cup E_{2})$. }

\begin{proof} See~\cite{CodishDY91}\end{proof}
\end{lemma}

%%%%%%%%%%

\proofoflemma{lm:scheme:vsldnf} { \label{pr:lm:scheme:vsldnf} VSLDNF
  and SLDNF deals with negative literals in the same manner.
  Therefore, it remains to prove for the cases where the leftmost
  literals are positive. The proof has two parts. The first part
  corresponds to procedure-entry and the second part to
  procedure-exit.

  We first prove that if $\sigma_{(j,k)}\uparrow
  \vcal_{i_j}\cong\rho_{j}\circ\tau_{(j,k)}\uparrow \vcal_{i_j}$ then
  R2 (p.\pageref{R2}) is derived from R1 (p.\pageref{R1}) iff R2'
  (p.\pageref{R21}) is derived from R1' (p.\pageref{R11}) and
  $\sigma_{(j+1,1)}\uparrow
  \vcal_{i_{j+1}}\cong\rho_{j+1}\circ\tau_{(j+1,1)}\uparrow
  \vcal_{i_{j+1}}$.

    Let $\sigma_{(j,k)}\vcal_{i_j}\cong\rho_{j}\circ\tau_{(j,k)}\uparrow
    \vcal_{i_j}$. Then there is a renaming $\delta$  such that
    \begin{equation}\label{loc:1} 
      (\sigma_{(j,k)}\uparrow
      \vcal_{i_j})\circ\delta=\rho_{j}\circ\tau_{(j,k)}\uparrow
      \vcal_{i_j}
    \end{equation}
    By choosing the same clause $C_{i_{j+1}}$ to be resolved with both
    R1 and R1', we have that R2 is derived from R1 iff R2' is derived
    from R1' by corollary~\ref{appendix:corollary:unification:3}
    (p.\pageref{appendix:corollary:unification:3}).  Suppose that R2
    were derived from R1 and R2' from R1'.

\begin{equation} 
\label{loc:2} 
\begin{array}[b] {lr}
B_{(i_j,k)}\rho_{j}\tau_{(j,k )} & \\
   ~= B_{(i_j,k)}(\rho_{j}\circ\tau_{(j,k)}) &\\ 
   ~= B_{(i_j,k)}(\rho_{j}\circ\tau_{(j,k)} \uparrow \vcal_{i_j}) & 
         (\because vars(B_{(i_j,k)})\subseteq \vcal_{i_j}) \\ 
   ~= B_{(i_j,k)}((\sigma_{(j,k)}\uparrow \vcal_{i_j})\circ\delta) &
   (\because equation~\ref{loc:1})\\ 
   ~=
  B_{(i_j,k)}\sigma_{(j,k)}\delta & (\because
  vars(B_{(i_j,k)})\subseteq \vcal_{i_j})
\end{array} 
\end{equation}

\begin{equation} \label{variant:loc1} 
\begin{array} [b]{lr}
\rho_{j+1}\circ\tau_{(j+1,1)}\uparrow \vcal_{i_{j+1}} &\\
  ~= \rho_{j+1}\circ\tau_{(j,k)}\circ\eta\uparrow \vcal_{i_{j+1}} 
  & (\because equation~\ref{loc:B})\\
   ~= \rho_{j+1}\circ\eta\uparrow \vcal_{i_{j+1}} 
  & (\because equation~\ref{loc:A})\\ 
  ~=\rho_{j+1}\circ
  mgu(H_{i_{j+1}}\rho_{{j+1}},B_{(i_j,k)}\rho_{j}\tau_{(j,k )})
  \uparrow \vcal_{i_{j+1}} 
  & (\because equation~\ref{loc:C}) \\
  ~=\rho_{j+1}\circ
  mgu(H_{i_{j+1}}\rho_{{j+1}},B_{(i_j,k)}\sigma_{(j,k)}\delta)
  \uparrow \vcal_{i_{j+1}}
  & (\because equation~\ref{loc:2})
\end{array} 
\end{equation}
Let $\overline{\delta}$ be the inverse of $\delta$. 
\begin{equation}     \label{variant:loc2} 
\begin{array} [b]{lr}
\sigma_{(j+1,1)} \uparrow \vcal_{i_{j+1}}&\\
 ~=mgu(H_{i_{j+1}},B_{(i_j,k)}\sigma_{(j,k)}\psi_{j+1})\uparrow\vcal_{i_{j+1}} 
   & (\because equation~\ref{loc:F})\\
  ~=mgu(H_{i_{j+1}},B_{(i_j,k)}\sigma_{(j,k)}\delta\overline{\delta}\psi_{j+1})\uparrow 
       \vcal_{i_{j+1}} & (\because \delta\overline{\delta}~is~identity)\\
 ~=mgu(H_{i_{j+1}},(B_{(i_j,k)}\sigma_{(j,k)}\delta)(\overline{\delta}\circ\psi_{j+1}))
      \uparrow\vcal_{i_{j+1}} & \\
 ~=mgu(H_{i_{j+1}},(B_{(i_j,k)}\sigma_{(j,k)}\delta)(\overline{\delta}\circ\psi_{j+1}))
      \uparrow vars(H_{i_{j+1}}) &
\end{array}
\end{equation}
$\sigma_{(j+1,1)}\uparrow \vcal_{i_{j+1}}\cong\rho_{j+1}\circ \tau_{(j+1,1)}\uparrow
\vcal_{i_{j+1}}$ by corollary~\ref{appendix:corollary:unification:3}
and
equations~\ref{variant:loc1}-\ref{variant:loc2}. This completes the
first part of the proof.

    We now prove that if
    $\sigma_{(j+1,m[i_{j+1}]+1)} \uparrow\vcal_{i_{j+1}}\cong\rho_{j+1}\circ\tau_{j,k+1}\uparrow
    \vcal_{i_{j+1}}$ then
    $\sigma_{(j,k+1)}\uparrow \vcal_{i_j}\cong\rho_{j}\circ\tau_{(j,k+1)}\uparrow
    \vcal_{i_j}$.  Let $\delta'$ be a renaming such
    that
    \[ \sigma_{(j+1,m[i_{j+1}]+1)}\uparrow\vcal_{i_{j+1}} =
    (\rho_{j+1}\circ\tau_{j,k+1}\uparrow
    \vcal_{i_{j+1}})\circ\delta'\] and $\overline{\delta'}$ be the
    inverse of $\delta'$.  
    \( \sigma_{(j+1,m[i_{j+1}]+1)} =
    \rho_{j+1}\circ\tau_{j,k+1}\circ\delta'\uparrow
     \vcal_{i_{j+1}}\). 
Therefore,
\begin{equation} \label{loc:3} 
\begin{array}[b]{lr}
{H_{i_{j+1}}\sigma_{(j+1,m[i_{j+1}]+1)}\phi_{j+1}} &\\
~= H_{i_{j+1}}\rho_{j+1}\tau_{(j,k+1)}\delta'\phi_{j+1} & \\
~= H_{i_{j+1}}\rho_{j+1}\tau_{(j,k)}\eta\theta\delta'\phi_{j+1} 
  &   (\because equation~\ref{loc:D})\\
~= H_{i_{j+1}}\rho_{j+1}\eta\theta\delta'\phi_{j+1} 
  & (\because vars(C_{i_{j+1}}\rho_{j+1})\cap vars(C_{i_{j}}\rho_{j})=\emptyset)\\
~= B_{(i_{j},k)}\rho_{j}\tau_{(j,k)}\eta\theta\delta'\phi_{j+1} &
( \because equation~\ref{loc:C})\\
\end{array} \end{equation}
By
equation~\ref{loc:1},
    \begin{equation} \label{loc:4} \sigma_{(j,k)}) \uparrow \vcal_{i_j} =
    \rho_{j}\circ\tau_{(j,k)}\circ\overline{\delta}\uparrow \vcal_{i_j}
    \end{equation} So, 
\begin{equation}\label{loc:5} 
\begin{array}[b]{lr}
B_{(i_j,k)}\sigma_{(j,k)} &\\
~=B_{(i_j,k)}(\rho_{j}\circ\tau_{j,k}\circ\overline{\delta}\uparrow\vcal_{i_j})
 & (\because equation~\ref{loc:4})\\
~=B_{(i_j,k)}\rho_{j}\tau_{j,k}\overline{\delta} 
 & (\because vars(B_{(i_j,k)})\subseteq \vcal_{i_j})
\end{array}\end{equation} 
Substituting equations~\ref{loc:3}-\ref{loc:5} into
equation~\ref{loc:G} and letting
$A=B_{(i_{j},k)}\rho_{j}\tau_{j,k}$, we have
\[
\begin{array}{lr}
{\sigma_{(j,k+1)}} \uparrow \vcal_{i_j} &  \\ 
~= \rho_{j}\circ\tau_{j,k}\circ\overline{\delta}\circ
  mgu(A\overline{\delta},A\eta\theta\delta'\phi_{j+1})
  \uparrow \vcal_{i_j} \\ 
~= \rho_{j}\circ\tau_{j,k}\circ(\overline{\delta}\circ
  mgu(A\overline{\delta},A\eta\theta\delta'\phi_{j+1})\uparrow
  vars(A) ) \uparrow \vcal_{i_j} & (\because
  lemma~\ref{appendix:lemma:composition:restriction}) \\
~\cong \rho_{j}\circ\tau_{j,k}\circ(mgu(A,A\eta\theta\delta'\phi_{j+1})\uparrow
  vars(A) )\uparrow \vcal_{i_j} 
  & (\because lemma~\ref{appendix:corollary:unification:2}) \\ 
~= \rho_{j}\circ\tau_{j,k}\circ mgu(A,A\eta\theta\delta'\phi_{j+1})\uparrow
  \vcal_{i_j} 
  & (\because lemma~\ref{appendix:lemma:composition:restriction})  \\
~= \rho_{j}\circ\tau_{j,k}\circ\eta\circ\theta\circ\delta'\circ\phi_{j+1}\uparrow
  \vcal_{i_j} & \\ 
~\cong \rho_{j}\circ\tau_{j,k}\circ\eta\circ\theta\uparrow\vcal_{i_j}
  & (\because \delta',\phi_{j+1}~are~renamings) \\ 
~= \rho_{j}\circ\tau_{(j,k+1)} \uparrow\vcal_{i_j} 
  & (\because equation~\ref{loc:D})
\end{array} 
\]
}

%%%%%%%%%%

\proofoflemma{lm:scheme:Fsharp}{\label{pr:lm:scheme:Fsharp} It is
  sufficient to prove that $\sem \uparrow k \subseteq
  \sgamma(\ssem \uparrow k)$ for any ordinal $k$.
  The proof is done by transfinite induction. 

  Basis.  $\sem \uparrow 0
  = \emptyset \subseteq \{\$\} = \sgamma(\sbot) =
  \sgamma(\ssem \uparrow 0)$.

  Induction. Let $\sem \uparrow k' \subseteq 
  \sgamma(\ssem \uparrow k')$ for any $k'<k$. If $k$
  is a limit ordinal then $\ssem \uparrow k =
  \ssqcup \{\ssem \uparrow k'~|~k'<k \}$. Therefore,
  $\sgamma(\ssem \uparrow k)\supseteq
  \sgamma(\ssem \uparrow k')$ for any $k'<k$ by
  equation~\ref{gammaint}. By the induction hypothesis,
  $\sgamma(\ssem \uparrow k)\supseteq \sem
  \uparrow k'$ for any $k'<k$. So, $\sem \uparrow k \subseteq
  \sgamma(\ssem \uparrow k)$. 

  Let $k$ not be a limit ordinal. Let
  $\stackitem{p_{1}}{q_{1}}{\theta_{1}}\cdots\stackitem{p_{n}}{q_{n}}{\theta_{n}}\cdot\$\in
  \sem \uparrow k$. By equation~\ref{gammaint}, it is sufficient
  to prove that $\edge{p_{i}}{q_{i}}\in\edges\wedge\theta_{i}\in
  {[\ssem \uparrow k]}_{\edge{p_{i}}{q_{i}}}$ for any
  $1\leq i\leq n$.

  There is $0\leq \jmath \leq 3$ such that
  $\stackitem{p_{1}}{q_{1}}{\theta_{1}}\cdots\stackitem{p_{n}}{q_{n}}{\theta_{n}}\cdot\$ \in
  \sem^{\jmath}(\sem \uparrow (k-1))$ by
  equation~\ref{eq:tsem}.

  Let $\jmath=0$. By equation~\ref{eq:tsem:2}, $n=1$ and
  $\edge{p_{1}}{q_{1}}\in \edges^{0}\wedge \theta_{1}\in\Theta_{p[1]}$.
  So, by equation~\ref{eq:csem:2},
  $\stackitem{p_{1}}{q_{1}}{\theta_{1}}\cdots\stackitem{p_{n}}{q_{n}}{\theta_{n}}\cdot\$
  \in \sgamma(\ssem\uparrow k)$.

  Let $\jmath=1$.  By equation~\ref{eq:tsem:1},
  $\edge{p_{1}}{q_{1}}\in\edges^{1}$, $p_{2}=q_{1}$ and
  $\stackitem{p_{2}}{q_{2}}{\theta_{2}}\cdots\stackitem{p_{n}}{q_{n}}{\theta_{n}}\cdot\$
  \in \sem  \uparrow (k-1)$ and $\theta_1=
  \unify(B_{q_{1}},\sigma,H_{p_{1}[1]},\epsilon)\neq \fail$.
  $\stackitem{p_{2}}{q_{2}}{\theta_{2}}\cdots\stackitem{p_{n}}{q_{n}}{\theta_{n}}\cdot\$
  \in\sgamma(\ssem\uparrow (k-1))$ by the induction
  hypothesis.
  $\stackitem{p_{1}}{q_{1}}{\theta_{1}}\cdots\stackitem{p_{n}}{q_{n}}{\theta_{n}}\cdot\$
  \in\sgamma( \ssem\uparrow k )$ by
  equation~\ref{eq:csem:1} and the monotonicity of $\ssem$.

Let $\jmath=2$.  By equation~\ref{eq:tsem:3}, $\edge{p_{1}}{q_{1}}\in\edges^{2}$
and there are two stack items
$\stackitem{q_{1}}{u}{\sigma}$ and $\stackitem{p_{1}^{\_}}{v}{\eta}$ such that
   \[
   \stackitem{q_{1}}{u}{\sigma}\cdot\stackitem{p_{1}^{\_}}{v}{\eta}\cdot \stackitem{p_{2}}{q_{2}}{\theta_{2}}\cdots\stackitem{p_{n}}{q_{n}}{\theta_{n}}\cdot\$ \in
       \sem \uparrow (k-1)
   \]
   \[ \theta_{1}=
   \unify(H_{q_{1}[1]},\sigma,B_{p_{1}^{\_}},\eta)\neq\fail\]
   $\stackitem{p_{2}}{q_{2}}{\theta_{2}}\cdots\stackitem{p_{n}}{q_{n}}{\theta_{n}}\cdot\$
   \in \sgamma(\ssem\uparrow (k-1))$ by the induction
   hypothesis. Therefore, 
  $\stackitem{p_{1}}{q_{1}}{\theta_{1}}\cdots\stackitem{p_{n}}{q_{n}}{\theta_{n}}\cdot\$
  \in\sgamma( \ssem\uparrow k )$ by 
   equation~\ref{eq:csem:3} and the monotonicity of $\ssem$.
 
   Let $\jmath=3$.  By equation~\ref{eq:tsem:4},
   $\edge{p_{1}}{q_{1}}\in\edges^{3}$, $p_{2}=q_{1}$, and
   $\stackitem{p_{2}}{q_{2}}{\theta_{2}}\cdots\stackitem{p_{n}}{q_{n}}{\theta_{n}}\cdot\$\in
   \sem \uparrow (k-1)$. By the induction hypothesis,
   $\stackitem{p_{2}}{q_{2}}{\theta_{2}}\cdots\stackitem{p_{n}}{q_{n}}{\theta_{n}}\cdot\$\in
   \sgamma(\ssem\uparrow (k-1))$. Therefore,
   $\stackitem{p_{1}}{q_{1}}{\theta_{1}}\cdots\stackitem{p_{n}}{q_{n}}{\theta_{n}}\cdot\$
   \in\sgamma( \ssem\uparrow k )$ by by
   equation~\ref{eq:csem:4} and  the monotonicity of $\ssem$.
 
   Therefore, $\sem \uparrow k \subseteq
  \sgamma(\ssem \uparrow k)$ for any ordinal $k$.
}

%%%%%%%%%%%%%

\proofoftheorem{th:safeness} {\label{pr:th:safeness} (C4) implies that
  $\fsem$ is monotonic and therefore $\lfp 
  \fsem$ exists.  By theorem~\ref{frame:th:cousot}, it
  suffices to prove that, for any $\fbigx\in \fdom$,
  \(\ssem \fcomp \fgamma(\fbigx) \ssqsubseteq
  \fgamma\fcomp \fsem(\fbigx)\).

  Let $\sigma\in {[\ssem \fcomp
    \fgamma(\fbigx)]}_{\edge{p}{q}}$. We need to prove $\sigma\in
  {[\fgamma\fcomp \fsem(\fbigx)]}_{\edge{p}{q}}$.  Let
  $\edge{p}{q}\in\edges^{\jmath}$ for some $0\leq\jmath\leq 3$.

 Let $\jmath=0$. By equation~\ref{eq:csem:2},
 $\sigma\in \asubgamma{\vcal_{p[1]}}(\ftheta_{p[1]})$.  By equation~\ref{eq:asem:2},
 $\sigma\in\asubgamma{\vcal_{p[1]}}({[\fsem(\fbigx)]}_{\edge{p}{q}}) =
 {[\fgamma\fcomp \fsem(\fbigx)]}_{\edge{p}{q}}$.

 Let $\jmath=1$. By equation~\ref{eq:csem:1}, there is $u\in \points$
 such that $\edge{q}{u}\in \edges$ and\linebreak \(\sigma\in
 \sunify(B_{q},{[\fgamma(\fbigx)]}_{\edge{q}{u}}, H_{p[1]},
 \{\epsilon\})\). By equations~\ref{eq:csem:1}
 and~\ref{e:concretisation}, C3 and the monotonicity of function
 $\sunify$ in its fourth argument,
\[\begin{array} {lll} \sigma&\in & \sunify(B_{q},
\asubgamma{\vcal_{q[1]}}(\fbigx_{\edge{q}{u}}),H_{p[1]},
\asubgamma{\vcal_{p[1]}}(\asubid{\vcal_{p[1]}}))\\ & \subseteq & \asubgamma{\vcal_{p[1]}}\fcomp \aunify{\vcal_{q[1]}}{\vcal_{p[1]}}(B_{q},\fbigx_{\edge{q}{u}}, H_{p[1]},
\asubid{\vcal_{p[1]}} )
\end{array}\]
So, by equations~\ref{e:concretisation} and~\ref{eq:asem:1} and the
monotonicity of $\asubgamma{\vcal_{p[1]}}$,
\[\begin{array} {lll} \sigma& \in &\asubgamma{\vcal_{p[1]}}
( {[ \fsem(\fbigx)]}_{\edge{p}{q}}) \\ & \subseteq &
\asubgamma{\vcal_{p[1]}} ( {[
  \fsem(\fbigx)]}_{\edge{p}{q}} )\\ & = & {[\fgamma\fcomp
  \fsem(\fbigx)]}_{\edge{p}{q}}
 \end{array} \]

 Let $\jmath=2$.  There are $u,v\in\points$
 such that  \mbox{$\edge{p^{\_}}{v}, \edge{q}{u}\in\edges$} and \[\sigma\in
 \sunify(H_{q[1]},{[\fgamma(\fbigx)]}_{\edge{q}{u}},B_{p^{\_}},
 {[\fgamma(\fbigx)]}_{\edge{p^{\_}}{v}}) \] by equation~\ref{eq:csem:3}. By
 equations~\ref{e:concretisation}
 and~\ref{eq:asem:3},
 
 \[ \begin{array} {lll} \sigma &\in &
 \sunify(H_{q[1]},\asubgamma{\vcal_{q[1]}}(\fbigx_{\edge{q}{u}}),B_{p^{\_}},
 \asubgamma{\vcal_{p^{\_}[1]}}(\fbigx_{\edge{p^{\_}}{v}}))
 \\ & \subseteq &
 \asubgamma{\vcal_{p^{\_}[1]}}\fcomp
 \aunify{\vcal_{q[1]}}{\vcal_{p[1]}}(H_{q[1]},\fbigx_{\edge{q}{u}}, B_{p^{\_}},
 \fbigx_{\edge{p^{\_}}{v}}) \\ & =
 &\asubgamma{\vcal_{p[1]}}\fcomp
 \aunify{\vcal_{q[1]}}{\vcal_{p[1]}}(H_{q[1]},\fbigx_{\edge{q}{u}}, B_{p^{\_}},
 \fbigx_{\edge{p^{\_}}{v}})\\ & \subseteq & \asubgamma{\vcal_{p[1]}} ( {[\fsem(\fbigx)]}_{\edge{p}{q}})\\ & = &
 {[\fgamma\fcomp \fsem(\fbigx)]}_{\edge{p}{q}}
 \end{array}\] 

 Let $\jmath=3$. By equation~\ref{eq:csem:4}, there is $u\in\points$
 such that \(\sigma\in {[\fgamma(\fbigx)]}_{\edge{q}{u}}\). By
 equations~\ref{e:concretisation} and~\ref{eq:asem:4},
 \[\begin{array}{lll} 
 \sigma &\in & {[\fgamma(\fbigx)]}_{\edge{q}{u}}\\ &=& \asubgamma{\vcal_{q[1]}}(\fbigx_{\edge{q}{u}})\\ 
 &\subseteq&\asubgamma{\vcal_{q[1]}}
 ({[\fsem(\fbigx)]}_{\edge{p}{q}})\\ & = & \asubgamma{\vcal_{p[1]}}({[\fsem(\fbigx)]}_{\edge{p}{q}})\\ 
 & = & {[\fgamma\fcomp \fsem(\fbigx)]}_{\edge{p}{q}} 
 \end{array} \]
}

%%%%%%%%%%

\proofoflemma{lm:safeness:simplified}{\label{pr:lm:safeness:simplified}
  It suffices to prove that, for any $\edge{p}{q}\in\edges$ and 
  any $\dbigx\in \ddom$, \(
  {[\fsem\fcomp\dgamma(\dbigx)]}_{\edge{p}{q}} = {[ \dgamma
    \dsem(\dbigx)]}_{\edge{p}{q}} \).  By
  equation~\ref{gamma:collapse},
  \begin{equation}\label{lm:safeness:simplified:loc} 
   \forall \edge{p}{q}\in{\edges}. (
   {[\dgamma(\dbigx)]}_{\edge{p}{q}} = \dbigx_{p} ) \end{equation}

   Let $\edge{p}{q}\in\edges^{0}$.  \( {[
     \fsem\fcomp\dgamma(\dbigx)]}_{\edge{p}{q}} =
   \ftheta_{p[1]} =
   {[\dgamma\fcomp \dsem(\dbigx)]}_{\edge{p}{q}} \) by
   equations~\ref{eq:asem:2}, \ref{sem:collapse:2}
   and~\ref{lm:safeness:simplified:loc}. 

   Let $\edge{p}{q}\in\edges^{1}$. By equations~\ref{eq:asem:1}, \ref{sem:collapse:1}
   and~\ref{lm:safeness:simplified:loc},
   \[ {[\fsem\fcomp\dgamma(\dbigx)]}_{\edge{p}{q}}
   = \aunify{\vcal_{q[1]}}{\vcal_{p[1]}}(B_{q}, \dbigx_{q},H_{p[1]}, \asubid{\vcal_{p[1]}}) =
   {[\dgamma\fcomp
     \dsem(\dbigx)]}_{\edge{p}{q}} \]
 
   Let $\edge{p}{q}\in\edges^{2}$. By
   equation~\ref{eq:asem:3}, \ref{sem:collapse:3}
   and~\ref{lm:safeness:simplified:loc},
  \[ {[\fsem\fcomp\dgamma(\dbigx)]}_{\edge{p}{q}}
  =
  \aunify{\vcal_{q[1]}}{\vcal_{p[1]}}(H_{q[1]},\dbigx_{q},B_{p^{\_}},\dbigx_{p^{\_}}) =
  {[\dgamma\fcomp \dsem(\dbigx)]}_{\edge{p}{q}}
  \]

   Let $\edge{p}{q}\in\edges^{3}$. By equations~\ref{eq:asem:4}, \ref{sem:collapse:4}
   and~\ref{lm:safeness:simplified:loc},
   \[
   {[\fsem\fcomp\dgamma(\dbigx)]}_{\edge{p}{p^{\_}}}  =
   \dbigx_{p^{\_}} =
  {[\dgamma\fcomp \dsem(\dbigx)]}_{\edge{p}{p^{\_}}}\]

  Therefore, \(
  {[\fsem\fcomp\dgamma(\dbigx)]}_{\edge{p}{q}} =
  {[ \dgamma \dsem(\dbigx)]}_{\edge{p}{q}} \).
  }

%\bibliography{referenc}
%\comments
{

}

\end{document}